\documentclass[aps,prd,showkeys,showpacs,amssymb,eqsecnum,amsfonts,epsf
preprintnumbers,nofootinbib,superscriptaddress]{revtex4}

\usepackage{graphicx}
\usepackage{natbib}
\usepackage{latexsym}
\usepackage{amsfonts}
\usepackage{url,hyperref}
\usepackage{bm}

\begin{document}

\begin{figure}[t]
\vspace{-1.2cm}
\hspace{-16.15cm}
\scalebox{0.09}{\includegraphics{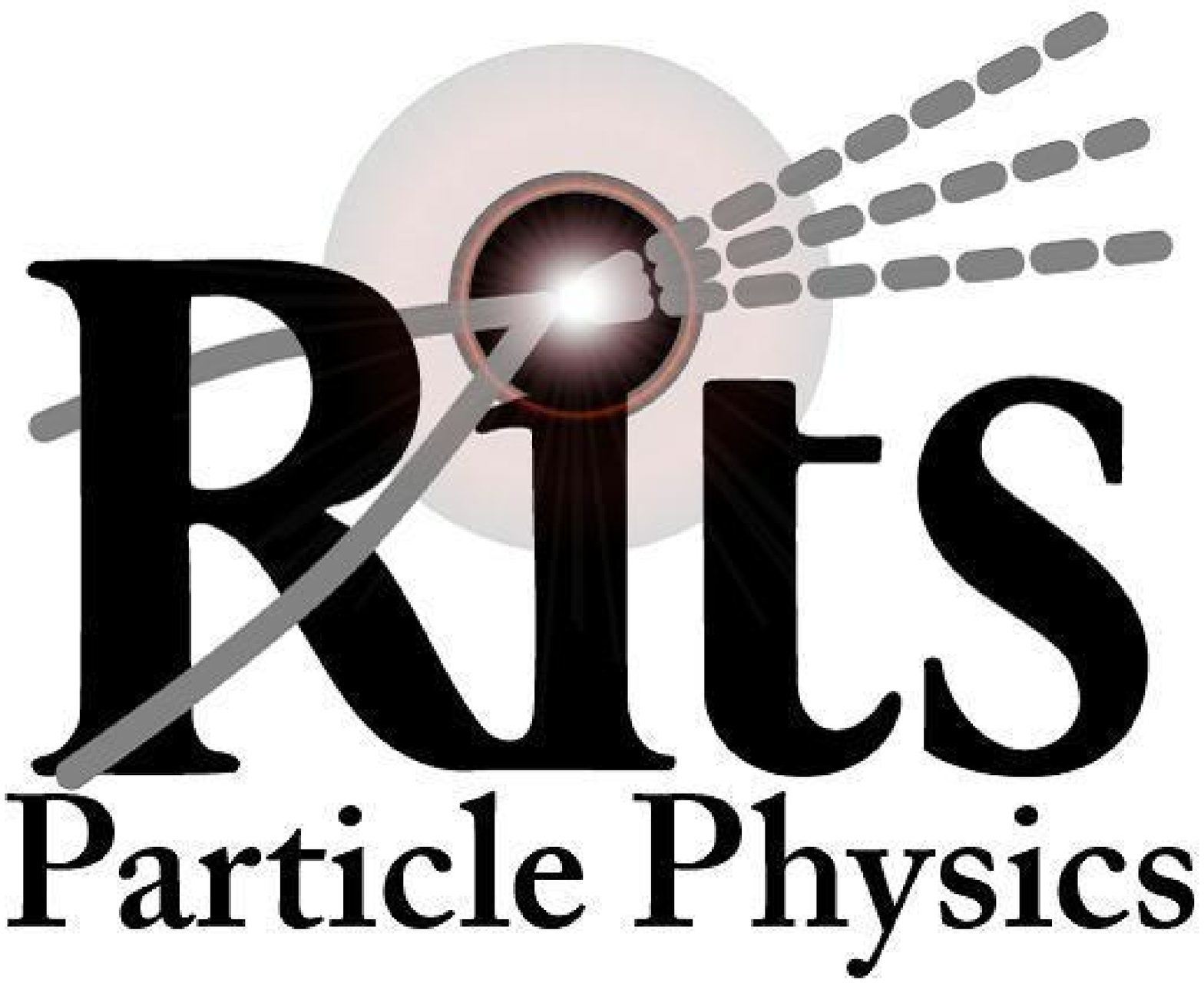}} 
\end{figure}

\newcommand{\vp}{\varphi}
\newcommand{\nn}{\nonumber\\}
\newcommand{\beq}{\begin{equation}}
\newcommand{\eeq}{\end{equation}}
\newcommand{\bed}{\begin{displaymath}}
\newcommand{\eed}{\end{displaymath}}
\def\bea{\begin{eqnarray}}
\def\eea{\end{eqnarray}}

\title{Volume stabilization in a warped flux compactification model}
\author{Masato~Minamitsuji}
\email[Email: ]{masato_AT_yukawa.kyoto-u.ac.jp}
\affiliation{Yukawa Institute for Theoretical Physics, Kyoto University, 
Kyoto 606-8502, Japan}
\author{Wade~Naylor}
\email[Email: ]{naylor_AT_se.ritsumei.ac.jp}
\affiliation{Department of Physics, Ritsumeikan University, 
Kusatsu, Shiga 525-8577, Japan}
\author{Misao~Sasaki}
\email[Email: ]{misao_AT_yukawa.kyoto-u.ac.jp}
\affiliation{Yukawa Institute for Theoretical Physics, Kyoto University, 
Kyoto 606-8502, Japan}

\begin{abstract}
We investigate the stability of the extra dimensions in a warped, codimension two braneworld that is based upon an Einstein-Maxwell-dilaton theory with
a non-vanishing scalar field potential. The braneworld solution has two 3-branes, which are located at the positions of the conical singularities.
For this type of brane solution the relative positions of the branes
(the shape modulus) are determined via the tension-deficit relations, if the brane tensions are fixed. However, the volume of the extra dimensions (the volume modulus) is not fixed in the context of the classical theory, implying we should take quantum corrections into account. Hence, we discuss the one-loop effective potential of the volume modulus for a massless, minimally coupled scalar field.  
Given the scale invariance of the background solution, the form of the modulus effective potential can {\it only} be determined from the sign of the logarithmic term in the effective potential that depends on the renormalization scale.
This term can be evaluated via heat kernel analysis and we show that in most cases the volume modulus is stabilized. In the actual evaluation, due to a lack of knowledge of the UV contributions from the conical branes in a six-dimensional spacetime, we consider its four-dimensional counterpart. We then go on to discuss the mass scale of the modulus itself and find that it becomes comparable to the gravitational scale for mild degrees of warping, when the renormalization scale is set nominally to the gravitational scale. Then, we make some suggestions on the original six-dimensional model. Finally, we close this article, after discussing some phenomenological implications relating to the hierarchy problem of the fundamental energy scales and the smallness of the effective vacuum energy on the brane. We find that one-loop corrections in this model appear to alleviate some of these problems.
\end{abstract}

\preprint{YITP-06-32}
\preprint{RITS-PP-010}
\date{\today}
\maketitle


\section{Introduction}
String theory suggests that our universe is not actually 
four-dimensional, but in fact a submanifold~(brane) embedded into a higher-dimensional spacetime~(bulk).
Braneworld gravity and cosmology, especially based on the proposal by
Randall and Sundrum \cite{Randall:1999ee,Randall:1999vf},
 has developed a lot in the literature in the codimension one context
 \cite{Langlois:2005nd}. String theory; however, 
suggests that there are as many as six or seven extra dimensions and 
thus, one may consider braneworld models with higher codimensions.
Recently, codimension two braneworld has been investigated eagerly, 
because it may give a more {\it natural} resolution of the cosmological constant problem than the codimension one case does, namely the vacuum energy of the brane may affect only the extra dimensions, not the geometry on the brane, see e.g., \cite{Carroll:2003db,Navarro:2003vw, Aghababaie:2003wz,Navarro:2003bf,Nilles:2003km, Vinet:2004bk,Garriga:2004tq,Burgess:2004dh,Vinet:2005dg,Tolley:2005nu}. 
The basic nature of gravity and cosmology in codimension two braneworld
has been reviewed, e.g., in
Ref. \cite{Papantonopoulos:2006uj} (see also references therein).

The stability of the extra dimensions is one of the most significant issues in
 braneworld models with two branes. The interbrane distance between the branes appears as
 a scalar degree of freedom in the four-dimensional effective theory and affects the geometry and cosmology on the brane. For instance, in RS-type codimension one brane models,  this modulus, called the radion, behaves as a scalar degree of freedom in an  effective Brans-Dicke gravity \cite{Garriga:1999yh}.
Several stabilization mechanisms have been discussed in the RS model, both 
by classical dynamics \cite{Goldberger:1999uk,Tanaka:2000er,Csaki:1999mp}
and quantum corrections to the bulk vacuum state
\cite{Garriga:2000jb,Garriga:2001ar}. Stability of the extra dimensions is also a significant issue in higher codimensions.
In this paper, we focus on the issue of quantum mechanical stabilization
of the modulus in codimension two braneworld.

Various codimension two braneworld solutions with two branes 
have been found, where the compact extra dimensions are supported by the dynamics of flux fields \cite{Carroll:2003db,Navarro:2003bf,Nilles:2003km,Vinet:2004bk,Garriga:2004tq,Vinet:2005dg}. Solutions with warped compact extra dimensions
have also been 
found \cite{Gibbons:2003di,Aghababaie:2003ar,Lee:2005az,Carter:2006uk}, especially in the context of six-dimensional supergravity theory 
\cite{Salam:1984cj,Nishino:1984gk,Nishino:1986dc}. 
In this paper, we focus on the solution discussed by Aghababaie et al. in
\cite{Aghababaie:2003ar}. Similar warped solutions have been recently discussed in \cite{Mukohyama:2005yw,Yoshiguchi:2005nn}, in the context of six-dimensional Einstein-Maxwell theory. In these two-brane models, 
the magnetic flux plays an essential role in order to obtain regular
warped solutions. By ``regular" we mean that singularities which are stronger than conical ones are not permitted.
The branes are situated at the conical defects, whose tensions are determined by 
the conical deficit angles. In warped brane models the conical defects correspond to the horizons of the bulk geometry and thus, the deficit angles depend on the 
parameters of the bulk geometry, e.g., the mass and charge.
By the tension-deficit relations, if the brane tensions are fixed, then part of bulk geometry is also fixed. In the models discussed in \cite{Mukohyama:2005yw,Yoshiguchi:2005nn}, the bulk geometry is also completely determined. However, in the model we shall consider here \cite{Aghababaie:2003ar},
 which is based on six-dimensional supergravity, only the warp factor is fixed by the
 classical analysis.  Thus, to fix the size of the extra dimensions, quantum corrections of the bulk field should be taken into account. In this paper, we consider such a supersymmetric warped flux compactification model.\footnote {Recently, scalar and tensor perturbations about supersymmetric codimension two brane models have been discussed, see e.g., \cite{Carter:2005hu,Carter:2006uk,Lee:2006ge,Peloso:2006cq}. }

We shall consider the perturbations of a massless scalar field on this brane
background and calculate the effective potential of the volume modulus, focusing on whether or not one-loop quantum corrections can stabilize the absolute size of the extra dimensions.
Due to the scale invariance of the background field theory, the form of
 the modulus effective potential is almost completely fixed and therefore we only need to evaluate the sign of the coefficient of the logarithmic term in the effective potential to investigate the possibility of volume stabilization. This is directly related to the appropriate heat kernel coefficient, which is composed of contributions from both the bulk and the conical branes.

The paper is organized as follows: In Sec. II, we present a warped flux compactification model that is based on an Einstein-Maxwell-dilaton theory with a non-vanishing scalar field potential.
In Sec. III,
we consider the perturbation of a massless, minimally coupled bulk scalar field in order to derive the modulus effective potential and then investigate whether or not the effective potential has a minimum.
As a result, the stability of the modulus is determined by the appropriate heat kernel coefficient. Then, we present results for our analysis of the heat kernel coefficients and we also discuss the {\it rigidity} of the stabilization, i.e., is the modulus mass stable against KK perturbations.
In the actual investigation of stability, rather than the original six-dimensional model, we consider a four-dimensional counterpart, because of the difficulties associated with evaluating  the UV contributions from the conical branes. Then, we make some suggestions for the original six-dimensional model. In Sec V, we discuss phenomenological implications relating to the hierarchy of the  fundamental energy scales and the effective vacuum energy realized on the brane. 
In Sec. VI, we summarize this paper and discuss possible extensions of our present analyses. 
In Appendix A, we introduce the four-dimensional version of the 
warped flux compactification solution and consider the nature of the massless scalar field perturbations on such a background, including quantum effects.
In Appendix B, we show the conformal invariance of the $a_{4}(f=1)$ heat kernel coefficient.
 In Appendix C we use the conformal invariance of the $a_{4,{\rm cone}}(f=1)$ heat kernel coefficient on the cone to find the smeared coefficient $a_{4,{\rm cone}}(f)$, which is required for the cocycle function.
In Appendix D we apply the WKB method to give an estimate for the zeta function in four dimensions, which contributes to the mass scale of the modulus.  


\section{A warped codimension two brane model with flux compactification}
 
 \subsection{Solution}
 
We consider a six-dimensional Einstein-Maxwell-dilaton theory with a non-vanishing scalar potential \cite{Burgess:2003mk} as
\begin{eqnarray}
  S_6=
M_6^4
\int d^6 x\sqrt{-g}
\left(
 \frac{1}{2}R
-\frac{1}{2}\partial_i \varphi \partial^i \varphi
-\frac{1}{4} e^{-\varphi}F_{ij}F^{ij}
 -2g_1^2 e^{\varphi} 
\right)\,,
\label{theory6d}
\end{eqnarray}
where $\varphi$ is a dilaton field, $F_{ij}$ represents a $U(1)$ gauge field strength and $g_1$ represents the dilaton potential.
This theory corresponds to the bosonic part of the Salam-Sezgin,
six-dimensional gauged supergravity
theory \cite{Salam:1984cj, Nishino:1984gk} with vanishing Kalb-Ramond 2-form and chiral scalar fields. 
Hereafter we set $M_6^4=1$ for simplicity and if needed, we
put it back in explicitly.  

The theory given in Eq. (\ref{theory6d}) has a dilatonic, charged black hole solution \cite{Burgess:2003mk} with metric
\begin{eqnarray}
ds^2&=& -h(\rho)dt^2 +\frac{d\rho^2}{h(\rho)}
     +(2\rho)(dx_1^2+dx_2^2+dx_3^2+dx_4^2)\,,
     \nonumber\\
     &&h(\rho)=-\frac{2\tilde M}{\rho}
               +\frac{\tilde A^2}{2\rho^3}
               -\frac{g_1^2\rho}{2}\,,
     \nonumber\\
     &&\varphi(\rho)=-\ln(2\rho)\,,
     \nonumber\\
     &&F_{t\rho}= \frac{\tilde A}{\rho^3}\,,
     \label{theory6dmet}
\end{eqnarray}
where $\tilde M$ and $\tilde A$ are integration constants. 

At this stage it will be convenient to perform a {\it double} Wick rotation, $t\to i\theta$, $x_1\to -i\tau$ with the reparameterizations $\tilde M\to -M$, $\tilde A\to i A$ \cite{Aghababaie:2003ar}. 
This leads to
\begin{eqnarray}
ds^2&= & h(\rho)d\theta^2
     +\frac{d\rho^2}{h(\rho)}
     +(2\rho)(-d\tau^2+d x_2^2+dx_3^2+dx_4^2)\,,
     \nonumber\\
     &&h(\rho)=\frac{2 M}{\rho}
              -\frac{A^2}{\rho^3}
              -\frac{g_1^2\rho}{2}\,,
     \nonumber\\
     &&\varphi(\rho)=-\ln(2\rho)\,,
     \nonumber\\
     &&F_{\theta\rho}= -\frac{A}{\rho^3}\,.
\label{bulkmetricori}
\end{eqnarray}
In this scenario, the branes are located at positions which are determined by
the {\it horizon} condition $h(\rho)=0$;
\begin{eqnarray}
 \rho_{\pm}^2
=\frac{2M}{g_1^2}
  \left(1\pm \sqrt{1-\frac{g_1^2 A^2}{4M^2}} \right)\,.
\end{eqnarray}
We can then rewrite
\begin{eqnarray}
h(\rho)=\frac{g_1^2}{2\rho^3}
       \left(\rho_+^2 -\rho^2\right)
       \left(\rho^2 -\rho_-^2\right)
\end{eqnarray}
and whence we also obtain the following useful relations
\begin{eqnarray}
\rho_+^2 + \rho_-^2 = \frac{4M}{g_1^2}\,,\qquad
\rho_+^2 \rho_-^2= \frac{A^2}{g_1^2}\,.
\label{parapara}
\end{eqnarray}
These give a direct relation between the mass and charge of the magnetic 
flux. The global period of $\theta$ is called $\Delta \theta$, which is
determined by the brane tensions through the tension-deficit relations as will be discussed later. Note that for the $\rho_-\to 0$ limit, we obtain a single brane solution with a naked curvature singularity in the bulk.


By changing coordinates to
\begin{eqnarray}
z= \left(\frac{\rho_+^2-\rho^2}{\rho^2-\rho_-^2}\right)^{1/2}
\label{mitsuo}
\end{eqnarray}
Eq. (\ref{bulkmetricori}) becomes
\begin{eqnarray}
ds^2= 2\rho
 \left[
 \frac{dz^2}{g_1^2 (1+z^2)^2}
+\frac{g_1^2 (\rho_+^2-\rho_-^2)^2}{4 (\rho_+^2+\rho_-^2 z^2)^2} z^2 d\theta^2
+\eta_{\mu\nu}dx^{\mu}dx^{\nu}
\right]\,.
\label{bulkmetric}
\end{eqnarray}
Thus, there are two cones at $z \to 0, \infty$,
whose deficit angles are given by
\begin{eqnarray}
\delta_+:=2\pi
         -\frac{g_1^2}{2}(1-r^2)\Delta\theta
\,,
\qquad\qquad
\delta_-:=2\pi -\frac{g_1^2}{2}\frac{1-r^2}{r^2}\Delta\theta\,,
\end{eqnarray}
respectively. Once the brane tensions are fixed, then $r=\rho_-/\rho_+$ is fixed.
But the absolute size of the extra dimensions, i.e., $\rho_+$,
is not fixed due to scale invariance of the bulk solution, unless
quantum corrections are taken into account. 
In the following, instead of fixing $\sigma_{\pm}$, we shall fix $\delta_+$ and $r$.


\subsection{Bulk geometry and branes}

The branes are codimension two boundaries, embedded onto conical deficits
 at $\rho=\rho_{\pm}$, whose actions are given by
\begin{eqnarray}
 S_{\pm}
=-\int d^4 x \sqrt{h}
 \sigma_{\pm}\,,
\end{eqnarray}
respectively, where $\sigma_{\pm}$ denotes the brane tensions. Note that 
we do not assume any coupling of the dilaton with the brane tensions and such vanishing dilaton couplings are required in order that effective cosmological constant on the branes vanishes completely, at least at the classical level \cite{Aghababaie:2003ar}. 
We also assume that there is no ordinary matter on the branes other than their tensions.
The brane tensions are related to the conical deficit angles by
\begin{eqnarray}
\sigma_{\pm}=  M_6^4 \delta_{\pm}\,.
\end{eqnarray}
We stress that these relations are only valid for sufficiently small brane tensions in comparison with the bulk scale $M_{6}^4$.
Then, the angle period $\Delta\theta$ is determined by
\begin{eqnarray}
 \Delta\theta
=\frac{2\pi-\delta_+}{\kappa_+}
=\frac{2\pi-\delta_-}{\kappa_-}\,, \label{angle period}
\end{eqnarray}
where $\kappa_{\pm}$ denotes the surface gravities on each brane:
\begin{eqnarray}
\kappa_{\pm}=\mp 
            \frac{1}{2}
            h'(\rho_{\pm})
           = \frac{g_1^2}{2\rho_{\pm}^2}
             (\rho_+^2 - \rho_-^2)\,.
\end{eqnarray}
Thus, we obtain
\begin{eqnarray}
 \frac{2\pi-\delta_+}{2\pi-\delta_-}
  =r^2\,. \label{tendef6d}
\end{eqnarray}
After eliminating $r$, we obtain
\begin{eqnarray}
 \Delta\theta
=\frac{2(2\pi-\delta_+)(2\pi-\delta_-)}
      {g_1^2(\delta_+-\delta_-)}\,.
\end{eqnarray}
We emphasize that, in general, $\Delta\theta \neq 2\pi$ implying that the bulk angular period 
can be completely determined by the brane tensions. For later convenience, we regard
it as a function of $r$ and $\delta_+$:
\begin{eqnarray}
\Delta\theta(r,\delta_{+})
=\frac{2(2\pi-\delta_+)}{g_1^2(1-r^2)}\,.
\end{eqnarray}

Once the brane tensions, $\sigma_+$ and $\sigma_-$ are fixed,
then $r$ is also fixed and thus, we now regard the free parameters as 
$r$ and $\delta_+$, along with the dilaton bulk coupling $g_1$.
The remaining degree of freedom used to determine the bulk geometry is 
the absolute size of the bulk, i.e., $\rho_+$. 
Due to the scale invariance of the background brane solution, $\rho_+$
can only be fixed by quantum corrections of bulk fields.
Note that there is also magnetic flux constraint given by
\begin{eqnarray}
 \int^{\rho_+}_{\rho_-} d\rho
 \int^{\Delta\theta}_0 d\theta
 F_{\rho\theta}
 = -\Delta\theta A
   \big(\frac{1}{\rho_-^2}-\frac{1}{\rho_+^2}\big)
 =-\frac{2(2\pi-\delta_+)}{g_1 r}  \,,
\end{eqnarray}
but the magnetic flux only depends on $r$ and $\delta_+$
and does not contribute to the volume of the bulk.

\subsection{The volume modulus}

We consider the modulus dynamics in a four-dimensional model in the context of the {\it moduli approximation}, assuming $\rho_+$ is a function of $x^{\mu}$.
This approximation may be valid for sufficiently low energy scales compared with
the bulk gravitational one $M_6$.
Like for the codimension one case, we expect that this approximation is valid
for low energies, namely when the energy scale on the brane is below that of
the bulk.  In this subsection, we shall reintroduce $M_6$ in the action in order to clarify the dimensionality.
We regard $\rho_+$ as a function of the coordinate on the brane
$x^{\mu}=(\tau, x_2, x_3, x_4)$ in
Eq.~(\ref{theory6dmet}) and thus, as the volume modulus.
Then, we expand the components of the Lagrangian density and take the modulus parts
out. It is rather useful to move to the four-dimensional physical frame:
\begin{eqnarray}
ds^2&=& 2\rho 
 \Big[
 \frac{dz^2}{g_1^2 (1+z^2)^2}
+\frac{g_1^2 (1-r^2)^2}{4 (1+r^2 z^2)^2} z^2 d\theta^2
+\eta_{\mu\nu}dx^{\mu}dx^{\nu}
\Big]
\nonumber\\
&=&2\rho_+(x^{\mu}) \tilde \rho(z) 
 \Big[
 \frac{dz^2}{g_1^2 (1+z^2)^2}
+\frac{g_1^2 (1-r^2)^2}{4 (1+r^2 z^2)^2} z^2 d\theta^2
+\eta_{\mu\nu}dx^{\mu}dx^{\nu}
\Big]
\nonumber\\
&=&2 \tilde \rho(z) 
 \Big[
 \rho_+(\tilde x^{\mu})\Big( \frac{dz^2}{g_1^2 (1+z^2)^2}
+\frac{g_1^2 (1-r^2)^2}
      {4 (1+r^2 z^2)^2} z^2 d\theta^2
\Big)
+\eta_{\mu\nu}d\tilde x^{\mu}d\tilde x^{\nu}
\Big]
\,,
\end{eqnarray}
where $\tilde x^{\mu}$ has the dimension of physical length.  
The metric form is just like a Kaluza-Klein theory with two compact dimensions, apart from the
overall conformal (warp) factor $2 \tilde \rho(z)$.

Then, we evaluate the effective action of the volume modulus 
\begin{eqnarray}
\frac{1}{2}
\sqrt{-g}
\big(R
    -(\partial \varphi)^2
\big)_{\rm mod}
=\frac{(1-r^2)z}{\rho_+ (1+z^2)^2}
\Big(
-(\tilde \partial \rho_+)^2
-4\rho_+ \tilde\partial^2 \rho_+
\Big)\,.
\end{eqnarray}
Note that the flux and dilaton potential terms do not give any kinetic term contributions to the moduli kinetic term because the vector potential $A_j$ has no scale dependence. 
The second term is a total derivative term and does not contribute
to the effective action. 
Integrating over the extra dimensions, we obtain
\begin{eqnarray}
 \big(S_6\big)_{\rm mod}
=\frac{M_6^4 (2\pi-\delta_+)}{2g_1^2}
 \int d^4 \tilde x 
 \left(-
  \frac{(\partial \rho_+)^2}{\rho_+}
  \right)
  \,.
\end{eqnarray}
After redefining the modulus as
\begin{eqnarray}
 \chi_6(x^{\mu})=
\sqrt{\frac{4M_{6}^4(2\pi-\delta_+)}{g_1^2}\rho_{+}}
\,,\label{effrad6d}
\end{eqnarray}
we obtain the canonical form of the modulus kinetic term as
\begin{eqnarray}
 \big( S_6 \big)_{\rm mod}
= \int d^4 \tilde{x}
 \Big(
 -\frac{1}{2}(\tilde \partial\chi_6)^2
 \Big)\,.
\end{eqnarray}

\section{The one-loop effective potential of the volume modulus}

Next, we introduce a massless, minimally coupled scalar field (not to be 
confused with the dilaton field that supports the background configuration).
From this we can investigate the one-loop effective action for such a scalar
field and thus, the effective potential of the volume modulus on the co-dimension two warped brane background.
From now on, we shall work mainly in Euclidean space, rather than 
in the original Lorentzian frame.

The action for the massless scalar field perturbations is given by
\begin{eqnarray}
S_{\rm scalar}=-\frac{1}{2}
                \int d^6 x\sqrt{g}
                \phi\Delta_6\phi\,. 
\label{scalaraction}
\end{eqnarray}

\subsection{Scalar one-loop effective action}

The one-loop effective action for a massless minimally coupled scalar field is defined as
\begin{eqnarray}
W_6= \frac{1}{2}{\rm ln }\,{\rm det} (-\Delta_6)\,,\label{effacc}
\end{eqnarray}
where $\Delta_6$ is the six-dimensional Laplacian, which is divergent and needs to be regularized and renormalized. For this purpose, we define
\begin{eqnarray}
   W_s
 =-\frac{\mu^{2s}}{2}
 \int^{\infty}_0 \frac{dt}{t^{1-s}}
  K(t,\Delta_6)\,,
\end{eqnarray}
where $K(t,\Delta_6)$ is the (integrated) heat kernel defined by
\begin{eqnarray}
 K(t,\Delta_6)
={\rm Tr}\left(e^{-t(-\Delta_6)}\right)\,.
\end{eqnarray}
The (integrated) zeta function is related to the heat kernel by 
a Mellin transform:
\begin{eqnarray}
 \zeta(s,\Delta_6)
=\frac{1}{\Gamma(s)}
\int^{\infty}_0 dt\,
t^{s-1} K(t,\Delta_6)
= {\rm Tr} \Big((-\Delta_6)^s \Big)
\label{zetafunc}
\end{eqnarray} 
and after analytically continuing to $s\to 0$ we obtain the renormalized one-loop effective action. The explicit expression for the renormalized effective action is
\begin{eqnarray}
  W_s
 = -\frac{1}{2} \mu^{2s}
   \Gamma(s)\zeta(s,\Delta_6)
 = -\frac{1}{2}\left(\frac{1}{s}
               -\gamma
               +\ln \mu^{2}
               \right)
          \zeta(0,\Delta_6)
      -\frac{1}{2}\zeta'(0,\Delta_6)
      +{\cal O}(s)\,,            
\end{eqnarray}
where $\gamma$ is Euler's constant $\gamma\approx 0.577216$ and the pole term at $s=0$ is removed by renormalization. 
Thus, after a redefinition of the renormalization scale we obtain the
renormalized effective action and the renormalized scalar field 
effective action can be written as
\begin{eqnarray}
W_{6, \rm ren}=-\frac{1}{2}\zeta'(0,\Delta_6)
              -\frac{1}{2}\zeta(0,\Delta_6)\ln \mu^2\,.
\end{eqnarray}
By integrating over the internal dimensions, the
four-dimensional effective potential is 
\begin{eqnarray}
W_{6,\rm ren}=\int \big(d^4 x \rho_+^2 \big) 
              V_{\rm 6, eff}
              =\int d^4{ \tilde x } 
              V_{\rm 6, eff}
               \,,\label{effact6d}
\end{eqnarray}
where $V_{\rm eff}$ has the dimensions $(length)^{-4}$. 
In the next section, we shall derive the effective potential in a codimension two warped brane model. For brevity, from now on we shall omit the subscript ``ren".

In the case of codimension two, the zeta function is given by the summation
\begin{eqnarray}
\zeta(s,\Delta_6)
           =\int d^4 x 
            \sum_{m,n}
           \int \frac{d^4k}{(2\pi)^4}
           \frac{1}{\lambda^{2s}}\label{zeta6d}
\end{eqnarray}
where the eigenvalues are defined by 
\begin{eqnarray}
 \Delta_6 \phi_{\lambda}
=-\lambda^2 \phi_{\lambda}\,.
\end{eqnarray}
It is straightforward to show that 
\begin{eqnarray}
 \zeta(0,\Delta_6)
 = a_6(f=1)\,,
\end{eqnarray}
where $a_6(f)$ is a heat kernel coefficient,
defined by the asymptotic expansion of the heat kernel 
\cite{Vassil,Burgess:2005cg}:
\begin{eqnarray}
K(t,\Delta_6)
\simeq 
\sum_{k\geq 0}
   t^{(k-6)/2}a_k(f)\,,
   \quad t\to 0\,.
\end{eqnarray}
In the warped codimension two case, the eigenvalues are unknown and 
 there are not even any analytic solutions for the eigenfunctions so 
we must resort to using an approximate WKB method to estimate the 
effective action.

\subsection{Continuous conformal transformations}

One strategy to evaluate the one-loop effective action 
and the effective potential is to define a continuous conformal transformation (parameterized by $\epsilon$)
\begin{eqnarray}
d{\tilde s}_{6,\epsilon}^2
 = e^{2(\epsilon-1)\omega}ds_6^2\,,\qquad
\omega= \frac{1}{2} \ln(2\rho)
\label{conformal}
\end{eqnarray}
and thus
\begin{eqnarray}
 d{\tilde s}_6^2
=(2\rho)^{\epsilon}
 \left(
   \frac{dz^2}{g_1^2(1+z^2)^2}
 +\frac{g_1^2(\rho_+^2 - \rho_-^2)^2 z^2}
          {4(\rho_+^2 + \rho_-^2 z^2)^2}
    d\theta^2
 + d{\bf x}^2
 \right)\,,
\label{conf}
\end{eqnarray}
where for $\epsilon=1$ we have the original metric, which we shall denote as
$\Delta_{6,\epsilon}=\Delta_6$. The classical action of this scalar field
is changed under a conformal transformation, see Eq. (\ref{conformal}),
by (here we are considering a massless, minimally coupled bulk scalar field)
\begin{eqnarray}
  S_{\rm scalar}
   =-\frac{1}{2}\int d^6 x\sqrt{g}
     \phi\Delta_6\phi
   =-\frac{1}{2}\int d^6 x\sqrt{\tilde g}
     \tilde \phi\Big(\tilde \Delta_6+E_6(\epsilon) \Big)
     \tilde \phi\,,
     \label{class6d}
\end{eqnarray}
where
\begin{eqnarray}
E_6(\epsilon)
&=& -4(\epsilon-1)^2 \tilde{g}^{ab}
                 \nabla_{a} \omega
                  \nabla_{b}\omega
     +2(\epsilon-1){\tilde \Delta}_6\ln\omega  
   \nonumber\\
 &=&
\left(\frac{1}{2\rho}\right)^{\epsilon}
\frac{g^2(1-\epsilon)(\rho_+^2 - \rho_-^2) 
   \big\{
     \rho_+^2 (2+(1-\epsilon)z^2)
+  \rho_-^2 z^2(-1+\epsilon-2z^2)
     \big\} }
     {(\rho_+^2 + \rho_-^2 z^2)^2}\,.
     \label{good}
\end{eqnarray}

Due to technical reasons, which will be explained later, we shall evaluate the zeta function in the non-warped frame $\epsilon=0$. The correction associated with such a conformal transformation is commonly known as the cocycle function (obtained from integration along the paths of the conformal transformation):
\begin{eqnarray}
  W_6
  &=&  
  -\frac{1}{2}
   \zeta'(0,\Delta_6)
  -\frac{1}{2}
   \zeta(0,\Delta_6)
   \ln \mu^2
 =
  -\frac{1}{2}
   \zeta'(0,\Delta_{6,\epsilon=0})
  -\frac{1}{2}
   \zeta(0,\Delta_{6,\epsilon=0})
   \ln \mu^2
  -\int_0^1
    d\epsilon\,
    a_6\, (f=\partial_{\epsilon}\ln \Omega_{\epsilon})\,.
\end{eqnarray}
Furthermore, thanks to the conformal invariance of $\zeta(0,\Delta_\epsilon)=a_6(f=1)$ we arrive at
\begin{eqnarray}
 W_{6}
& =&  -\frac{1}{2}
   \zeta'(0,\Delta_{6,\epsilon=0})
  -\frac{1}{2}
   \zeta(0,\Delta_{6,\epsilon=0})
   \ln \mu^2
  -\int_0^1
    d\epsilon\,
    a_6\, (f=\partial_{\epsilon} (\epsilon-1)\omega)
\nonumber\\
&=& -\frac{1}{2}
   \zeta'(0,\Delta_{6,\epsilon=0})
  -\frac{1}{2}
   a_6(f=1)
   \ln \mu^2
  -\int_0^1
    d\epsilon\,
    a_6\, (f=\frac{1}{2}\ln(2\rho))
\nonumber\\
&=&   -\frac{1}{2}
          a_6(f=1)
            \ln (\mu^2\rho_+)
+\Big\{
-\frac{1}{2}
   \zeta'(0,\Delta_{6,\epsilon=0})
  -\int_0^1
    d\epsilon\,
   \Big(
   a_6 (f=\frac{1}{2}\ln(2\rho))
  -\frac{1}{2}\ln(\rho_+)
   a_6(f=1)
  \Big)
\Big\}
\nonumber\\
&=&
  -\frac{1}{2}
          a_6(f=1)
            \ln (\mu^2\rho_+)
+\Big\{
-\frac{1}{2}
   \zeta'(0,\Delta_{6,\epsilon=0})
  -\int_0^1
    d\epsilon\,
   a_6 (f=\frac{1}{2}\ln(\frac{2\rho}{\rho_+}))
\Big\}\,.\label{dec}
\end{eqnarray}
The term $a_6$ is given by the volume integration of linear combinations of cubic order curvature tensors \cite{Vassil,Burgess:2005cg,Hoover:2005uf}:
\begin{eqnarray}
  a_6(f)
&:=& (4\pi)^{-3}\Bigl\{
\int_M d^6x \sqrt{g}\,
 \Big[ \frac f{7!}\big(
      18R^{;i}{}_{;i}{}^{;j}{}_{;j}
    +17R_{;k}R^{;k}
    -2R_{ij;k}R^{ij;k} 
    -4R_{jk;n}R^{jn;k}
    +9R_{ij kl;n}R^{ij kl;n}
    +28RR_{;n}{}^{;n}
    \nonumber \\
&&\qquad
    -8R^{jk}R_{jk;n}{}^{;n} 
    +24R^{jk}R_{j}{}^{n}{}_{;kn}
    +12R_{ij kl}R^{ij kl ;n}{}_{;n}
    +\frac{35}{9}R^{3}
    -\frac{14}{3}RR_{ij}R^{ij} 
    +\frac{14}{3}R R_{ijkl}R^{ijkl}
       \nonumber \\
&&\qquad
    -\frac{208}{9}R_{jk}R^{j}{}_{n}R^{kn}
       +\frac{64}{3}R_{ ij}R_{kl}R^{ik jl}
      -\frac{16}{3}R_{jk}R^{j}{}_{n l i}R^{kn l i}
     +\frac{44}{9}R_{ij kn}R^{ij}{}_{lp}R^{kn l p} 
     +\frac{80}{9}R_{ij kn}R^{i}{}_{l}{}^{k}{}_{p}
          R^{jl  np}
   \big)
       \nonumber \\
&&\qquad
     +\frac{f}{360} 
     \big(
      6E_6{}^{;i}{}_{;i}{}^{;j}{}_{;j}
     +60E_6E_{6}{}^{;i}{}_{;i}
     +30E_{6}^{;i}E_{6;i}
     +60E_{6}^{3}
     +10R E_{6}{}^{;k}{}_{;k}
     +4R^{jk}E_{6;jk}
     +12R^{;k}E_{6;k}
     \nonumber \\
&&\qquad
     +30E_6^2R
     +12E_6R^{;k}{}_{;k}
     +5E_6R^2
     -2E_6R^{ij}R_{ij}
     +2E_6R^{ijkl}R_{ijkl}
     \big)   
 \Big]
     \nonumber\\
 &&
 \qquad
 +({\rm contribution\,\, of\,\, conical\,\, branes})
 \Big\}
   \,.\label{a6}
\end{eqnarray}


\subsection{Effective potential of the volume modulus}

In order to avoid a volume divergence 
we shall employ the one-loop effective potential, as given by Eq. (\ref{effact6d}), to 
try and stabilize the extra dimensions.
From our discussion in the previous subsection, the effective potential
takes exclusively the following form
\begin{eqnarray}
V_{\rm 6, eff}(r,\delta_+, \rho_+,\mu)
=\frac{A_6(r,\delta_+)-B_6(r,\delta_+) \ln(\mu^2 \rho_+)}{\rho_+^2}\,,
\label{originaleffpot}
\end{eqnarray}
where we have defined
\begin{eqnarray}
 \int d^4 x A_6(r,\delta_+)
 &=& \int d^4 \tilde x \frac{ A_6(r,\delta_+)}{\rho_+^2}
  = -\int_0^1 
       d\epsilon 
       a_6(f=\frac{1}{2}\ln(\frac{2\rho}{\rho_+})) 
  -\frac{1}{2}\zeta'(0,\Delta_{6,\epsilon=0})
    \,,
      \nonumber\\
\int d^4 x B_6(r,\delta_+)
 &=&\int d^4 \tilde x \frac{B_6(r,\delta_+)}{\rho_+^2}
  =\frac{1}{2}\zeta(0,\Delta_{6,\epsilon=0})
  =\frac{1}{2} a_6 (f=1)\,.  \label{6dcoef}
\end{eqnarray}
 Clearly, if $B_6(r,\delta_+)>0$,
 then the modulus effective potential has a minimum at
\begin{eqnarray}
 \rho_{+}^\ast
=\mu^{-2}  e^{(2A_6+ B_6)/(2B_6)}\,.\label{sta-san}
\end{eqnarray}
After a redefinition of the modulus, as given by Eq. (\ref{effrad6d}),
the effective potential can be rewritten as
\begin{eqnarray}
 V_{\rm 6,eff}(r,\delta_+,\chi,\mu):
=\Big(
\frac{4M_6^4 (2\pi-\delta_+)}{g^2}
 \Big)^2
\frac{A_6(r,\delta_+)
     -B_6(r,\delta_+)
       \ln
       \big(
       \frac{\mu^2 g^2\chi_6^2}
            {4M_6^4(2\pi-\delta_+)}
       \big)}
     {\chi_6^4}\,. 
\end{eqnarray}
The field value at the minimum is then given by
\begin{eqnarray}
\chi_{6, \ast}^2
=\frac{4M_6^4(2\pi-\delta_+)}
      {\mu^2 g^2}
e^{(2A_6+B_6)/(2B_6)}\,\label{effmodu}
\end{eqnarray}
and hence, the effective modulus mass becomes
\begin{eqnarray}
   m_{\rm 6, eff}^2
:=\partial_{\chi_6}^2 V_{\rm 6,eff} \Big|_{\ast}
= \frac{8g_1^2}{2\pi-\delta_+}
            B_6(r,\delta_+)
            e^{-3(2A_6+B_6)/(2B_6)}
             \Big(\frac{\mu}{M_6}\Big)^6
             M_6^2
             \,.
\label{6dmass}
\end{eqnarray}
Thus, the ratio between the stabilized mass and the effective bulk cosmological
constant is
\begin{eqnarray}
\frac{m_{\rm 6, eff}^2}{g_1^2 \rho_{+}^{-1}}
=\frac{8}{2\pi-\delta_+}
 B_6(r,\delta_+)
 \Big(\frac{\mu}{M_6}\Big)^4
  e^{-(2A_6+B_6)/B_6}\,. \label{conp6d}
\end{eqnarray}
The expression above allows us to quantify the {\it rigidity} of stabilization; namely, is 
the modulus mass stable against KK perturbations. 
However, the problem is that the $a_6$ contribution for conical branes has never 
been formulated as far as the authors are aware.


\section{Volume stabilization}

In this section, in order investigate the one-loop effective potential of the volume modulus, we 
consider the four-dimensional counterpart of the six-dimensional case
in the theory \cite{Burgess:2003mk},
\begin{eqnarray}
  S_4
=M_4^2 
\int d^4 x\sqrt{-g}
\left(
  R
-\frac{1}{2}\partial_i \varphi \partial^i \varphi
-\frac{1}{8} e^{-\varphi}F_{ij}F^{ij}
 -4g^2 e^{\varphi} 
\right)\,,\label{theory}
\end{eqnarray}
where, again, $\varphi$ is a dilaton, $F_{ij}$ represents a $U(1)$ gauge field strength and $g$ represents the dilaton potential (corresponding to $g_1$ in
the original model).
Hereafter, we set the bulk gravitational scale to $M_4^2=1$ for simplicity;
we shall reinsert it if and when it is needed.
The essential properties of this solution is summarized in Appendix A.
The reason we use this model is that we can evaluate the UV contributions from the conical branes. 
We shall extrapolate our results to the original model in six dimensions.

\subsection{One-loop effective potential of the volume modulus}

The zeta function and one-loop effective action in four dimensions
can be defined similarly to the case of six dimensions, as in
Eq. (\ref{zeta6d}) and Eq. (\ref{effact6d}), just by replacing 
 $``6" \to ``4"$ (and $(2\pi)^4\to (2\pi)^2$).
 Thus, the one-loop effective potential of the volume modulus takes the same form as Eq. (\ref{effact6d}), after
the replacement of $d^4 \tilde x$ with $d^2\tilde x$. 
 The effective action also takes the same form as Eq. (\ref{dec}), given 
the scale invariant nature of the flux compactification solution, implying the
effective potential can be written similarly as
\begin{eqnarray}
V_{\rm 4, eff}(r,\delta_+, \rho_+,\mu)
=\frac{A_4(r,\delta_+)-B_4(r,\delta_+) \ln(\mu^2 \rho_+)}{\rho_+}\,,
\end{eqnarray}
where 
\begin{eqnarray}
 \int d^2 x A_4(r,\delta_+)
 &=& \int d^2 \tilde x \frac{ A_4(r,\delta_+)}{\rho_+}
 =  -\int_0^1 d\epsilon 
       a_4(f=\frac{1}{2}\ln(\frac{2\rho}{\rho_+}))
  -\frac{1}{2}\zeta'(0,\Delta_{4,\epsilon=0})
\,,
      \nonumber\\
\int d^2 x B_4(r,\delta_+)
 &=&\int d^2 \tilde x \frac{B_4(r,\delta_+)}{\rho_+} 
  =\frac{1}{2}\zeta(0,\Delta_{4,\epsilon=0})
  =\frac{1}{2} a_4 (f=1)\,.  \label{4dcoef}
\end{eqnarray}
Note that the stabilized volume of the internal space is given by
\begin{eqnarray}
(\rho_{+,\ast})^{1/2}
= \mu^{-1} e^{(A_4+B_4)/(2B_4)}\,.
\label{exponent}
\end{eqnarray}

Again, if $B_4(r,\delta_+)>0$,
 then the modulus effective potential has a minimum at $\rho_+=\rho_{+,\ast}$,
 which is determined by $A_4(r,\delta_+)$ and $B_4(r,\delta_+)$.
Importantly, given the similarity of the spacetime structure in 
this four dimensional model with the original six dimensional model,  
the form of the effective modulus field
can be estimated by dimensional arguments, in comparison to Eq. (\ref{effmodu}). Namely,
\begin{eqnarray}
 \chi_4 (x^{\mu})
:=\sqrt{\frac{M_4^2(2\pi-\delta_+)}{g^2}
 \rho_+ }
\label{effrad4d}
\end{eqnarray}
acts as the modulus field in two dimensions.
The field value at the minimum is then given by
\begin{eqnarray}
\chi_{4, \ast}^2
=\frac{M_4^2(2\pi-\delta_+)}
      {\mu^2 g^2}
e^{(A_4+B_4)/B_4}\,
\end{eqnarray}
and hence, the effective modulus mass is
\begin{eqnarray}
m_{\rm 4, eff}^2:=\partial_{\chi_4}^2
              V_{\rm 4, eff} \Big|_{\ast}
             = \frac{4g^2}
                    {2\pi-\delta_+}
           B_4(r,\delta_+)
            e^{-2(A_4+B_4)/B_4}
             \Big(\frac{\mu}{M_4}\Big)^4
             M_4^2
             \,.
\label{4dmass}
\end{eqnarray}
The {\it rigidity} of the stability is determined by the quantity
\begin{eqnarray}
\frac{m_{\rm eff}^2 }{g^2 (\rho_{+,\ast})^{-1}}
= \frac{4}{2\pi-\delta_+}
           B_4(r,\delta_+)
            e^{-(A_4+B_4)/B_4}
             \Big(\frac{\mu}{M_4}\Big)^2\,. 
             \label{massratio}
\end{eqnarray}

Given the conformal invariance of the heat kernel coefficients, $a_4(f)$ is straightforward to evaluate in the $\epsilon=1$ frame.
In the original four-dimensional spacetime, including the contribution of the cones \cite{Dowker:1994bj,Fursaev:1994pq,Mann:1996bi} we have
\begin{eqnarray}
  a_4(f=1)
&:=&(4\pi)^{-2}
  \Big\{360^{-1}
   \int_M d^4 x \sqrt{g}
       \Big(
             12R^{;k}{}_{;k}
            +5R^2 
            -2R_{ij}R^{ij}
            +2R_{ijkl}R^{ijkl}
       \Big)
\nonumber\\
  &-&\int d^2 x \sum_{A=\pm}
  \left(\frac{\delta_A}{2\pi}\right) 
  \frac{2-\frac{\delta_A}{2\pi}}
       {1-\frac{\delta_A}{2\pi}}
  \sqrt{h_A}
\Big[
  \frac{\pi}{3}
  \Big(    
  \frac{1}{6}R
  +\lambda_1 \sum_a \big(
  \kappa^{(a)}{}^2
     -2 \kappa^{(a)}_{ij}\kappa_{(a)}^{ij}
                  \big)
  \Big)
     \nonumber
     \\ 
  &+&\frac{\pi}{180}
  \frac{ 
     2-\frac{\delta_A}{\pi}
     +\frac{\delta_A^2}{4\pi^2} }
     {(
     1-\frac{\delta_A}{2\pi})^2}
  \Big(\sum_a R_{aa}
       -2\sum_{a,b} R_{abab}
   +\frac{1}{2}\sum_a \kappa^{(a)}{}^2
   +\lambda_2\sum_a
   \big(\kappa^{(a)}{}^2
     -2 \kappa^{(a)}_{ij}\kappa_{(a)}^{ij}
   \big)
   \Big)
 \Big]
   \Big\},\label{a4}
\end{eqnarray}
where $\{a \}$ runs from $1$ to $2$ which denotes the two orthonormal directions to the conical branes
\begin{eqnarray}
\sum_{a,b} R_{abab}&=&
         \sum_{a,b} R_{ijkl}
           n^{i}_{(a)}n^{j}_{(b)}n^{k}_{(a)}n^{l}_{(b)}\,,
\quad
 \sum_a R_{aa}
 =\sum_a R_{ij}n^i_{(a)} 
              n^j_{(a)}\,
 \label{conecurve}
\end{eqnarray}
and $\kappa^{(a)}_{ij}$ is the extrinsic curvature defined by
 $\kappa^{(a)}_{ij}= -h_{i}^k h_j^l  \nabla_k n^{(a)}_l$. Here $n^{(a)}_i$ denotes two mutually orthogonal (inward pointing) unit normals and $a= z,\, \theta$ and $\kappa^{(a)} = h^{ij}\kappa^{(a)}_{ij}$, where $h_{ij}=g_{ij}-\sum_a n^{(a)}_i n^{(a)}_j$ is the induced metric
on the brane. The terms $\lambda_{1,2}$ are undetermined parameters associated with conformal
invariance of the conical heat kernels. 
 But, as shown in Appendix C, the extrinsic curvatures vanish on the cones
 and we need not specify them.
Note that this formula is valid for any deficit angle
\cite{Dowker:1994bj,Mann:1996bi}.

\subsection{Stability of the volume modulus in four dimensions}


Now we present the results of our
analyses on the one-loop effective potential of the modulus.
The function $B_4(r,\delta_+)$ that corresponds to $B_6(r,\delta_+)$ in the original six-dimensional model
is composed of two parts; a bulk and conical brane contribution and thus, the total 
is
\begin{eqnarray}
  B_4(r,\delta_+)
= \big(B_4(r,\delta_+)\big)_{\rm bulk}
+ \big(B_4(r,\delta_+)\big)_{\rm branes}.
\end{eqnarray}
From the properties of the heat kernel coefficients $B_4(r,\delta_+)$
is invariant under conformal transformations, as is explicitly 
demonstrated in Appendix C.
The bulk part of $B_4(r,\delta_+)$ is obtained from 
Eq. (\ref{bulk}) for $\epsilon=1$ and as shown in Appendix B, it is in fact independent
of $\epsilon$.
Similarly, the brane contribution to $B_4(r,\delta_+)$ is obtained from 
Eq. (\ref{inflationary}).

In Fig.~1, the bulk contribution to $B_4(r,\delta_+)$ is shown as a function of $x$, for $g=1$ and $\delta_+=0.01$.
Note that in the following plots, which are described as functions of $r$, 
we have set $\delta_+=0.01$.  There is no significant dependence on $\delta_+$ and thus we mainly focus on $r$.
For other values of $g$, the behavior of $B_4(r,\delta_+)$ is also almost the same.
The bulk contribution is always positive and as $r\to 0$ it suffers from a divergence because of the presence of a naked curvature singularity.
\begin{figure}
\begin{center}
  \begin{minipage}[t]{.45\textwidth}
   \begin{center}
    \includegraphics[scale=.8]{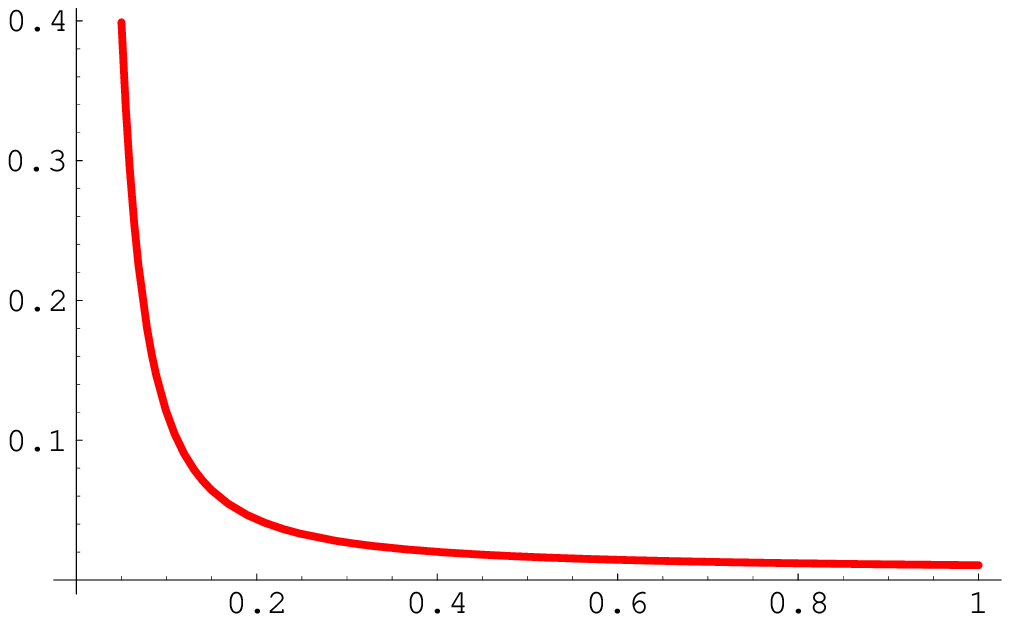}
        \caption{
  The bulk contribution to $B_4(r,0.01)$ is shown as a function of $r$, for $g=1$.}  
   \end{center}
   \end{minipage} 
\hspace{0.5cm}
   \begin{minipage}[t]{.45\textwidth}
   \begin{center}
    \includegraphics[scale=.8]{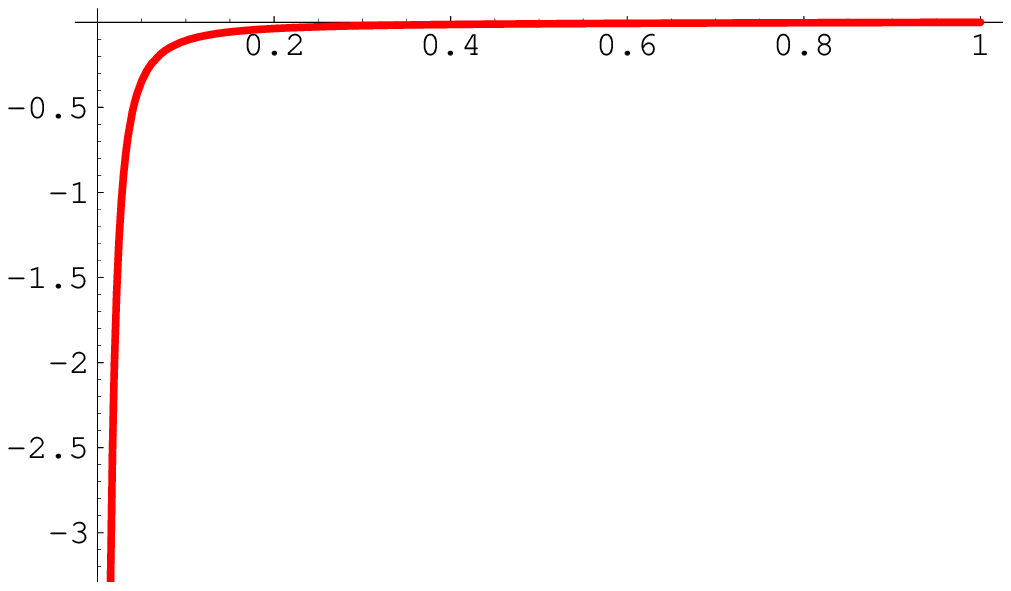}
\caption{
Contribution of conical branes to $B_4(r,0.01)$ is shown as a function of $r$, for $g=1$. 
       }          
   \end{center}
   \end{minipage}
   \end{center}
\end{figure}
In Fig. 2, the contribution from the branes, to $B_4(r,\delta_+)$,
is shown as a function of $r$.
 Comparing this with Fig. 1, one can readily see 
that the cone contribution becomes comparable to that of the bulk one but
the sign is opposite. 
Thus, the positivity of the total, $B_4(r,\delta_+)$, is a rather non-trivial issue.
In Fig. 3, the total, $B_4(r,\delta_+)$, is shown as a function of $r$ and
we find that the sign of $B_4(r,\delta_+)$ is always positive. 
Thus, for all values of $r$, the modulus can be stabilized;
this is one of the main results in this article. 
Though here we discuss the four-dimensional counterpart,
we will make suggestions on the more realistic case of six dimensions later.
Note that the behavior as shown in Fig. 3 is almost the same
for other values of the conical deficit angle $\delta_+$.

\begin{figure}
\begin{center}
  \begin{minipage}[t]{.45\textwidth}
   \begin{center}
    \includegraphics[scale=.8]{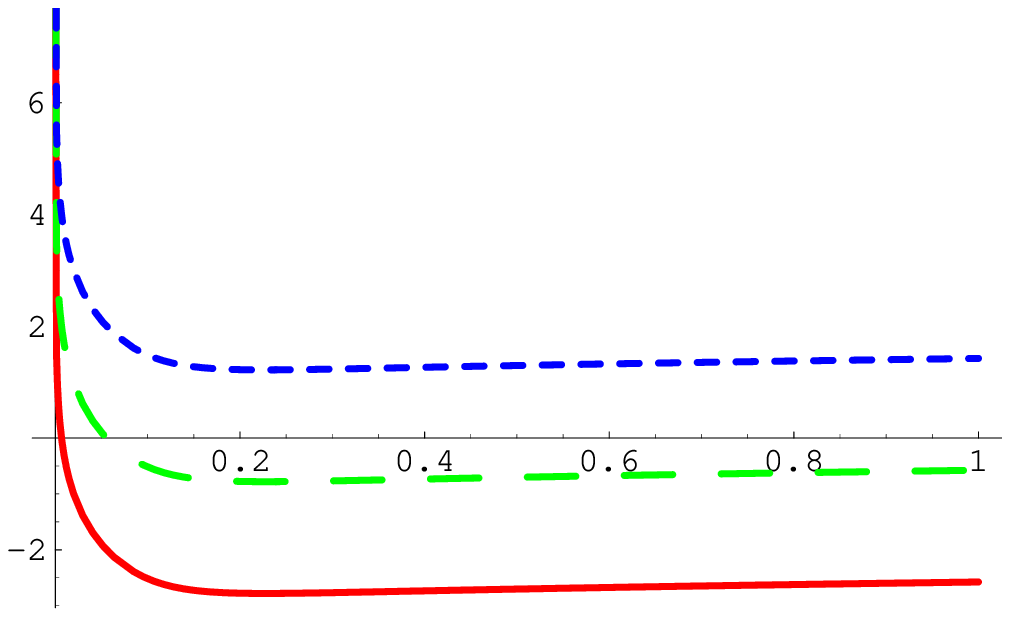}
 \caption{The total $\log_{10}(B_4(r,0.01))$ is shown. 
The red (solid), green (dashed, with wider intervals) and blue (dashed, with shorter intervals) curves correspond to $g=0.5,~5,~50$, respectively.
          }  
   \end{center}
   \end{minipage} 
\hspace{0.5cm}
   \begin{minipage}[t]{.45\textwidth}
   \begin{center}
    \includegraphics[scale=.8]{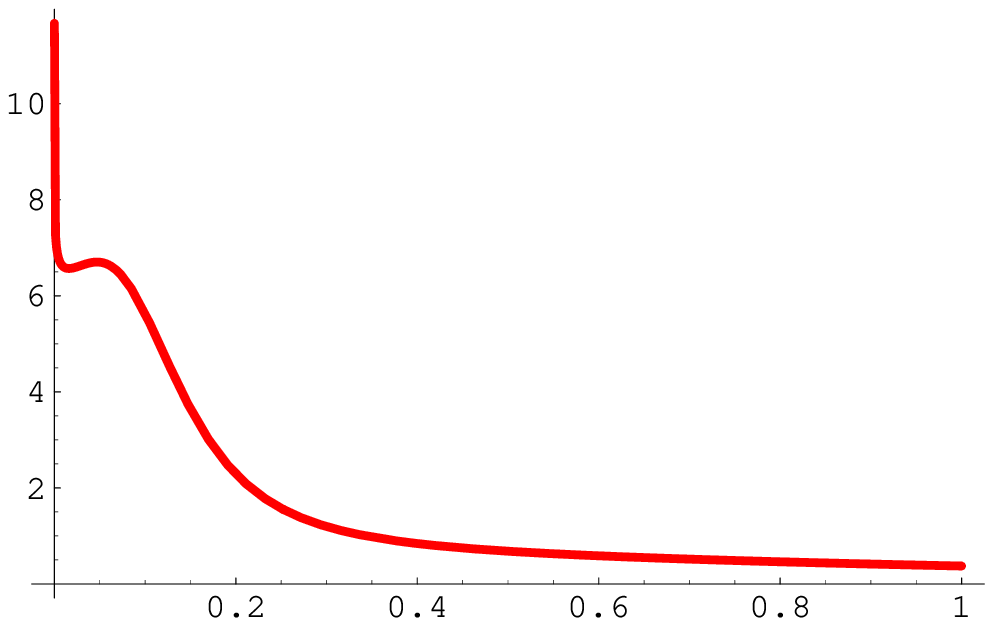}
\caption{ A logarithmic plot of the exponential factor in Eq. (\ref{exponent}) is shown as a function of $r$ for $g=1$ and $\delta_+=0.01$. In the following  
all logarithmic plots are to the base $10$.
       }          
   \end{center}
   \end{minipage}
   \end{center}
\end{figure}

 Next, we wish to discuss the minimum potential energy and and thus, the effective mass of the modulus. These can only be considered through the combination
 $A_4(r,\delta_+)/B_4(r,\delta_+)$, see Eq. (\ref{exponent}), where
 $A_4(r,\delta_+)$ corresponds to $A_6(r,\delta_+)$ in the original model.
 This ratio should then be compared with the renormalization scale, $\mu$.
We give an explicit derivation of the cocycle function and the derivative of the zeta function in Appendix C and D, respectively.

The term $A_4(r,\delta_+)$ is obtained from Eq. (\ref{4dcoef}).
The first term is the derivative of the zeta function in
the non-warped conformal frame, $\epsilon=0$, in the context of WKB approximation. The second term is known as the cocycle 
function, which corresponds to corrections associated with a conformal
transformation. By combining these two contributions we obtain
\begin{eqnarray}
  A_4(r,\delta_+)
= \big(A_4(r,\delta_+)\big)_{\rm cocycle}
+ \big(A_4(r,\delta_+)\big)_{\rm WKB}. \label{a_tot}
\end{eqnarray}
The bulk and brane contribution to the cocycle function, 
$A_{4, {\rm cocycle}}(r,\delta_+)$, is given in Eq. (\ref{inflationary2}), while
the WKB contribution, $A_{4, {\rm WKB}}(r,\delta_+)$, in the non-warped frame,
is given in Eq. (\ref{inflationary3}).
Actually, the value of $A_4(r,\delta_+)$ itself is not as important as the ratio
$A_{4}(r,\delta_+)/B_4(r,\delta_+)$, because the stabilized size of the extra dimensions 
is given by Eq. (\ref{exponent}).
In Fig. 4,~we show a logarithmic plot of the exponential factor as a function of $r$ for $g=1$ and $\delta_+=0.01$.
For all values of $r$, the stabilized size of the extra dimensions is larger than that of the renormalization scale. Especially in the limits $r\to 0$ and $r\to 1$, much larger extra  dimensions are permitted. 

Furthermore, in order to show how the stability of the extra dimensions are set {\it rigid}, we can consider the ratio between the effective modulus mass and the bulk curvature scales $g^2 (\rho_{+,\ast})^{-1}$.
In Fig. 5, we show logarithmic plots of Eq.~(\ref{massratio}) as a function of
$r$ for various dilaton couplings $g$.
\begin{figure}
\begin{center}
  \begin{minipage}[t]{.45\textwidth}
   \begin{center}
    \includegraphics[scale=.8]{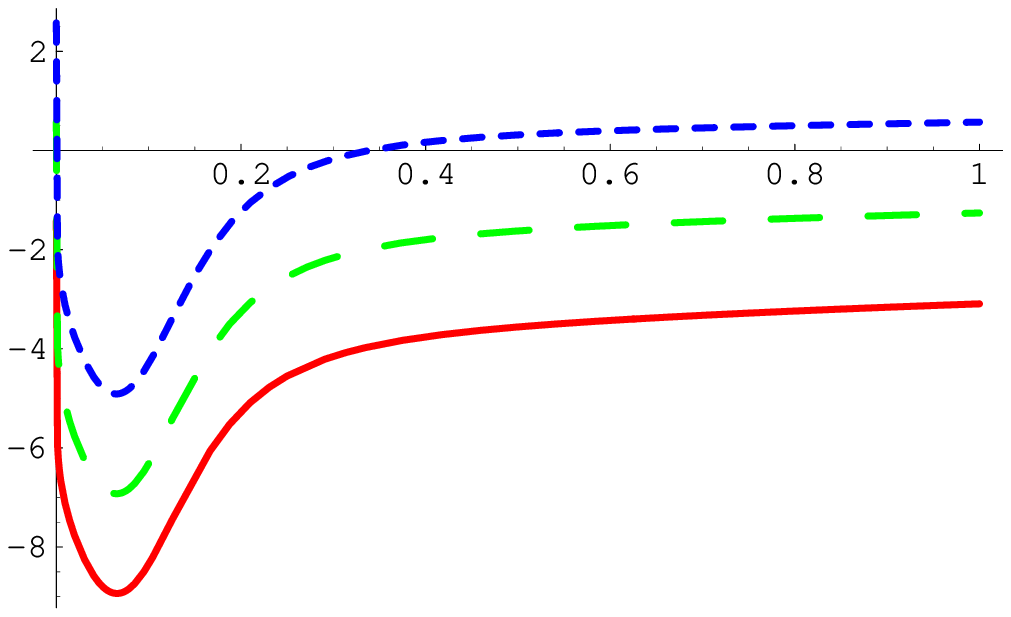}
        \caption{
  A logarithmic plot of Eq~(\ref{massratio}) for $\mu=M_4$
   is shown as a function of $r$. 
The red (solid), green (dashed, with wider intervals) and blue (dashed, with shorter intervals) curves correspond to $g=0.5,~5,~50$, respectively and
$\delta_+=0.01$.
        }  
   \end{center}
   \end{minipage} 
\hspace{0.5cm}
   \begin{minipage}[t]{.45\textwidth}
   \begin{center}
    \includegraphics[scale=.8]{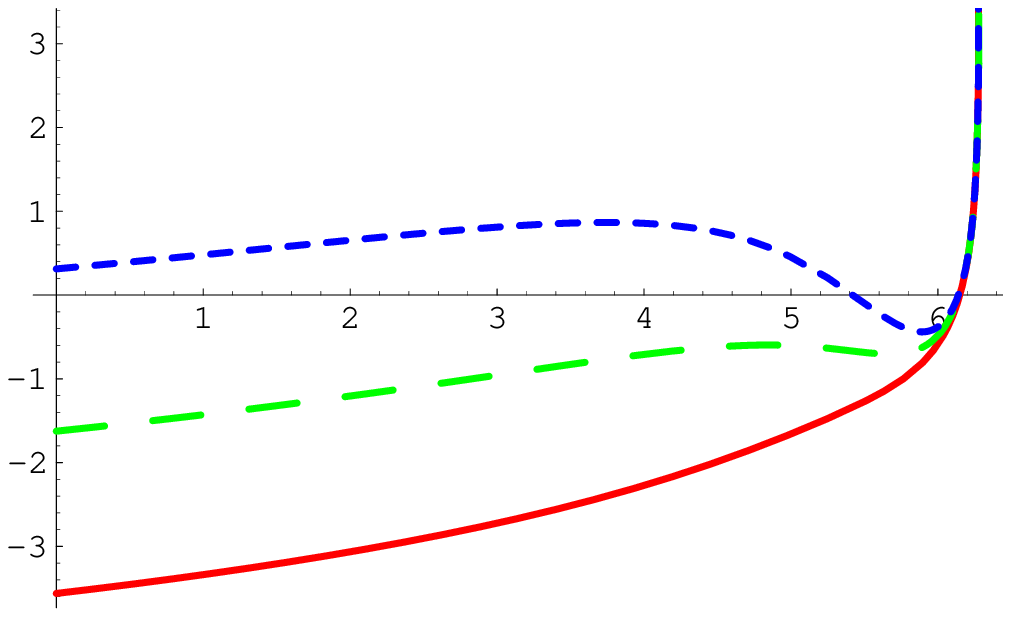}
\caption{  A logarithmic plot of Eq~(\ref{massratio}) for $\mu=M_4$
   is shown as a function of $\delta_+$. 
The red (solid), green (dashed, with wider intervals) and blue
(dashed, with shorter intervals) curves correspond to $g=0.5,~5,~50$,
respectively and
$r=0.5$.
       }          
   \end{center}
   \end{minipage}
   \end{center}
\end{figure}
In Fig. 6, we also show them as a function of $\delta_+$.
If the ratio is greater than unity, then, the stability of the modulus is generic.
From Fig.~5 and 6, the value of Eq. (\ref{massratio}) becomes larger for larger couplings $g$ (in fact the relative mass is almost proportional to $g^2$).
Thus, for large dilaton couplings the stability of the modulus is rigid against KK perturbations.
In Fig. 5, we also see that for smaller values of $r$, the modulus mass becomes
relatively light, whereas for larger $r$ it is insensitive to $r$.
In the limit $r\to 0$ the modulus mass diverges, because a bulk naked singularity is formed.
As a final comment, Fig. 6 is not particularly sensitive to changes in $\delta_+$, but it does exhibit a divergent behavior for $\delta_+ \to 2\pi$. This is because in such a limit the bulk essentially disappears.

\subsection{Suggestions for volume stabilization in six dimensions}

Now, we shall make some suggestions on
the stability of the volume modulus in six dimensions.
Because of our poor knowledge about the contribution from the conical branes 
we can evaluate only the bulk contribution. However, we can compare the result of the bulk contribution in six-dimensions to the four-dimensional one discussed previously.

Evaluating the bulk part of $B_6(r,\delta_+)$ from Eqs. 
(\ref{a6}) and (\ref{6dcoef}) leads to
\begin{eqnarray}
\big(B_6(r,\delta_+)\big)_{\rm bulk}
 &=&\frac{1}{2}
    \int^{\infty}_0 dz
    \frac{1}{64\pi^2}
   \frac{g_1^4( 2\pi-\delta_+)z}{2(1+z^2)(1+r^2 z^2)}
   \frac{1}{1260}
   \frac{\Psi(1,r,z )}
   {(1+r^2 z^2)^6}\,, \label{b6conc}
\end{eqnarray}
where
\begin{eqnarray}
\Psi(\rho_+,\rho_-,z)
&:=&
2035\,\rho_+^{12}{z}^{6}
+1564\, \rho_-^{4}\rho_+^{8}
-548\,\rho_-^{6}\rho_+^{6}
+5461\, \rho_-^{12}{z}^{8}
+4964\, \rho_-^{12}{z}^{10}
+2035\, \rho_-^{12}{z}^{6}
+1748\, \rho_-^{12}{z}^{12}
\nonumber\\
&
+&5461\, \rho_+^{12}{z}^{4}
-1740\, \rho_+^{10} \rho_-^{2}
+4964\, \rho_+^{12}{z}^{2}
+1748\, \rho_+^{12}
+32534\,\rho_+^{8} \rho_-^{4}{z}^{4}
+26109\,\rho_+^{8}{z}^{6}\rho_-^{4}
+13944\, \rho_+^{8}\rho_-^{4}{z}^{2}
\nonumber\\
&+&12757\,\rho_+^{8}{z}^{8} \rho_-^{4}
-2882\,\rho_+^{10}{z}^{6} \rho_-^{2}
-9049\, \rho_+^{10}{z}^{4}\rho_-^{2}
-5936\, \rho_+^{10} \rho_-^{2}{z}^{2}
-548\,\rho_+^{6}{z}^{12} \rho_-^{6}
+1564\,{z}^{12} \rho_+^{4}\rho_-^{8}
\nonumber\\
&-&1740\,{z}^{12} \rho_+^{2} \rho_-^{10}
+711\, \rho_+^{10}{z}^{8}\rho_-^{2}
+3652\, \rho_+^{8}{z}^{10} \rho_-^{4}
-9049\,{z}^{8} \rho_+^{2} \rho_-^{10}
-2882\,\rho_-^{10}\rho_+^{2}{z}^{6}
+711\, \rho_-^{10} \rho_+^{2}{z}^{4}
\nonumber\\
&-&5936\,{z}^{10} \rho_+^{2} \rho_-^{10}
+32534\,{z}^{8}\rho_+^{4}\rho_-^{8}
+13944\,{z}^{10} \rho_+^{4} \rho_-^{8}
+26109\, \rho_+^{4} \rho_-^{8}{z}^{6}
+3652\, \rho_-^{8} \rho_+^{4}{z}^{2}
+12757\, \rho_-^{8} \rho_+^{4}{z}^{4}
\nonumber\\
&-&10480\, \rho_-^{6} \rho_+^{6}{z}^{2}
-30044\, \rho_+^{6} \rho_-^{6}{z}^{6}
-27054\,\rho_+^{6}{z}^{8} \rho_-^{6}
-10480\, \rho_+^{6} \rho_-^{6}{z}^{10}
-27054\, \rho_+^{6} \rho_-^{6}{z}^{4}\,.
\end{eqnarray}


In Fig.~7,
we plot Eq. (\ref{b6conc}) as a function of $r$ for $g_1=1$ and
$\delta_+=0.01$ (this behavior is essentially the same for other values of the coupling $g_1$).
The behavior of the bulk part of $B_6(r,\delta_+)$
is similar to that of
$B_4(r,\delta_+)$ as shown in Fig. 1. As we mentioned previously, we have no analytic formulae to evaluate the contribution from the conical branes
for $a_6(f)$.
However, by analogy we expect similar behavior for the conical part of the heat kernel coefficient.
It may be worth mentioning that for larger $r$ and appropriate values of $g_1$, $B_6(r,\delta_+)$ is very likely to be positive and thus the modulus can probably be stabilized.
We could also evaluate $\zeta'(0,\Delta_6)$, via the WKB method, and due to symmetry would follow  almost identical steps to the four dimensional case, see Appendix C. However, we have no way to include the cocycle correction, at the level of our semi-analytic approach.
\begin{figure}
\begin{center}
  \begin{minipage}[t]{.45\textwidth}
   \begin{center}
    \includegraphics[scale=.8]{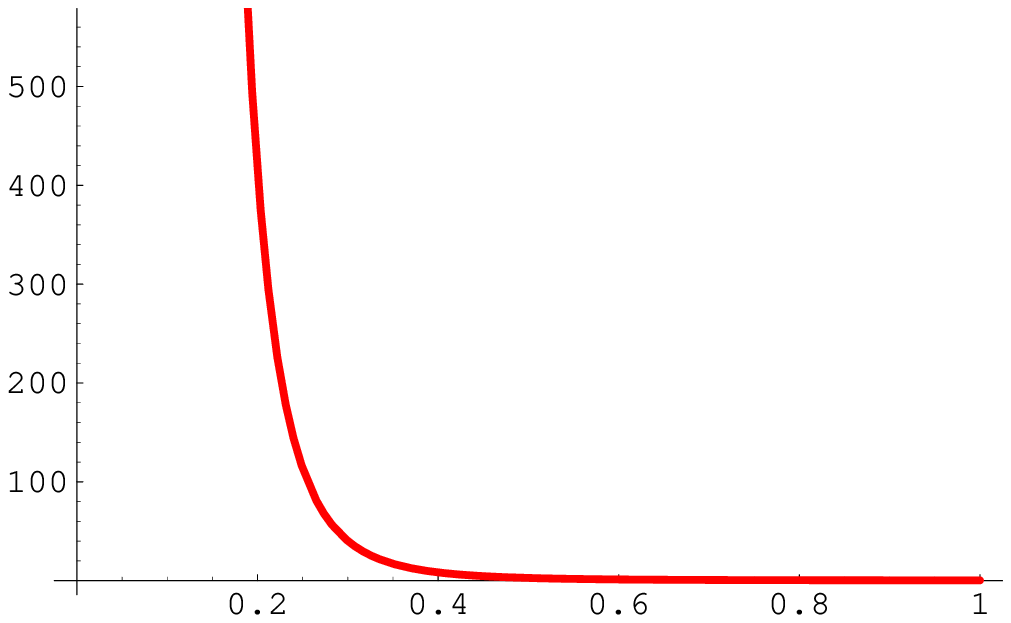}
\caption{
The bulk part of $B_6(r,0.01)$ is shown as a function of $r$ for $g_1=1$.}  
   \end{center}
   \end{minipage} 
\hspace{0.5cm}
   \begin{minipage}[t]{.45\textwidth}
   \begin{center}
    \includegraphics[scale=.8]{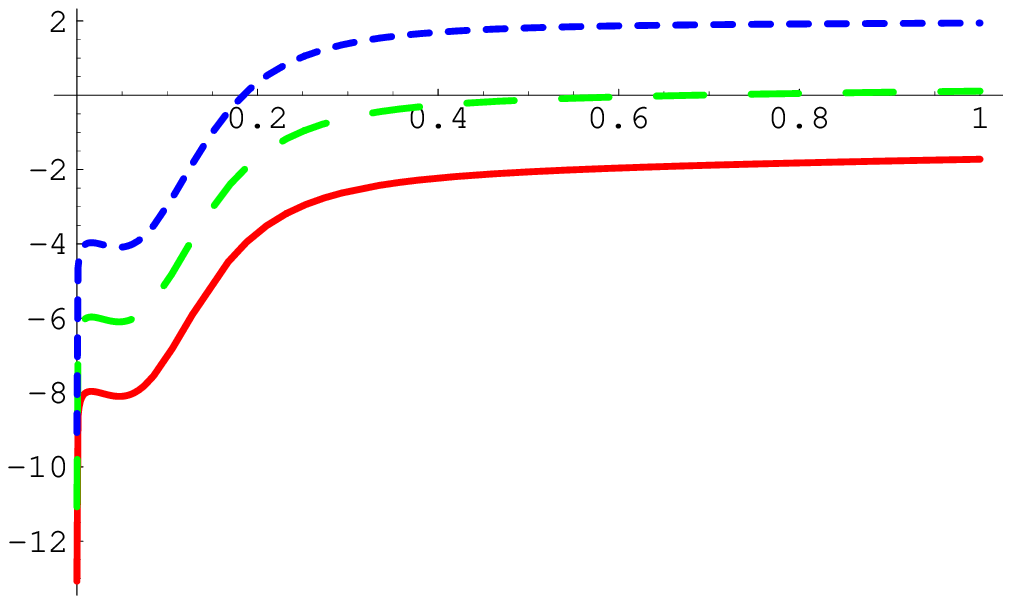}
\caption{ $\log_{10} (R(r, 0.01))$
is shown as a function of $r$, for $\delta_+=0.01$.
The red (solid), green (dashed, with wider intervals) and blue
(dashed, with shorter intervals) curves correspond to $g=0.5,~5,~50$,
respectively. }          
   \end{center}
   \end{minipage}
   \end{center}
\end{figure}

Thus, any further investigation of the conical contribution and the effective mass of the stabilized modulus (in six dimensions) is somewhat out of the scope of this article and is left for future work.

\section{Phenomenological Implications}

Finally, before closing this article, we shall discuss some phenomenological implications, i.e.,
the hierarchy structure between the fundamental energy scales
and the effective vacuum energy on the brane. Even though we are investigating, 
 primarily, the four-dimensional case, we expect similar results to hold in six dimensions.

\subsection{Hierarchy of the fundamental scales}

One of the most longstanding issues in theoretical physics to date, is the problem of the hierarchy structure of the fundamental energy scales, i.e., the huge gap between the gravitational scale, $M_{\rm pl}\sim 10^{19}{\rm  GeV}$, and the electroweak scale, $M_{\rm EW} \sim 10^{3} {\rm GeV}$.

In the original six-dimensional model, the {\it effective} four-dimensional Planck 
scale\footnote{Not to be confused with $M_4$ in the four-dimensional model.} is
 \begin{eqnarray}
   M_{\rm pl}^2
 \simeq
 \frac{\rho_+(2\pi-\delta_+)}{g_1^2}
  M_{6}^4\,.     
 \end{eqnarray}
 If we assume a brane localized field whose bare mass is given by $m^2$ 
 on either brane at $\rho_{\pm}$ then the observed mass scales are
 \begin{eqnarray}
 m_{+}^2= m^2\,,\qquad \qquad
 m_{-}^2= r^2 m^2\,.
 \end{eqnarray}
Thus, the mass ratio between the field and the effective Planck mass is given by
\begin{eqnarray}
\frac{m_{+}^2}{M_{\rm pl}^2}
\simeq
\Big(  \frac{\mu^2 m^2}{M_{6}^4}\Big)
  \frac{g_1^2}{2\pi-\delta_+}
   e^{-(2A_6 +  B_6)/(2B_6)}\,,
   \qquad\qquad
 \frac{m_{-}^2}{M_{\rm pl}^2}
\simeq
\Big(  \frac{\mu^2 m^2}{M_{6}^4}\Big)
  r^2
  \frac{g_1^2}{2\pi-\delta_+}
   e^{-(2A_6+B_6)/(2B_6)}\,,
  \end{eqnarray}
Assuming that the factor of $(\mu m/M_{6}^2)^2$ takes the optimal value of ${\cal O}(1)$
at the unification of the fundamental scales, the mass ratio becomes
\begin{eqnarray}
  \frac{g_1^2}{2\pi-\delta_+}
   e^{-(2A_6 + B_6)/(2B_6)}\,,
\end{eqnarray}
where we have used the value of $\rho_{+,\ast}$, given by Eq. (\ref{sta-san}),
at stabilization.
At present, the best we can do is use the results obtained from the four-dimensional model:
\begin{eqnarray}
R(r,\delta_+)
 = \frac{g^2}{2\pi-\delta_+}
   e^{-(A_4+B_4)/B_4}\,.\label{hier}
\end{eqnarray}
The results are shown in Fig. 8 and indicate that smaller values of $g$ and $r$ 
lead to a large hierarchy on both branes. However, 
the mass of the stabilized modulus is perhaps too light 
to support the realization of the hierarchy at the fundamental scales.
This is somewhat similar to the case of RS codimension one branes
\cite{Garriga:2000jb,Garriga:2001ar}.

\subsection{Effective vacuum energy density}

Another interesting issue is the origin of the observed vacuum energy density,
which is much smaller than the field theoretical prediction; the so-called the cosmological constant problem. Given the background brane solution that we are considering, the brane vacuum energy (the brane tensions) only contributes to the bulk geometry and therefore does not act as
an energy source for today's accelerated cosmic expansion
(at the background level the branes are to be kept flat). 
However, in this model, the observed vacuum energy density can arise from
one-loop quantum corrections as follows.

At the stabilized volume, from Eqs. (\ref{originaleffpot}) and (\ref{sta-san}),
the effective potential takes the value
\begin{eqnarray}
V_{\rm 6, eff}^{\ast}=-\frac{1}{2}
         \mu^4 B_{6}(r,\delta_+)  e^{-(2A_6+B_6)/ (2A_6)}
\end{eqnarray} 
and hence, the brane realized vacuum energy is almost completely determined by the renormalization
scale. It is then quite natural that the renormalization scale is taken to the unification  scale, i.e., of order $M_6$. However, in such a case the exponential factor is also very small 
and therefore we may be able to obtain a sufficiently small vacuum energy, which acts as a dark energy source. Although the vacuum energy is negative definite one might expect some kind of uplifting mechanism, for instance, due to the presence of test branes \cite{Kachru:2003aw}.

As in the case of the hierarchy problem, we shall consider the potential energy at the time of the stabilization. Whence, the vacuum energy is given by
\begin{eqnarray}
V_{\rm 4, eff}^{\ast}=-
         \mu^2 B_{4}(r,\delta_+)  e^{-(A_4+B_4)/ A_4}\,.
\end{eqnarray}
The coefficient acting on the renormalization scale is very similar to
the quantity plotted in Fig. 5.
Again, for smaller $g$ and $r$, we can expect smaller
vacuum energy densities of the volume modulus.
The mass of the modulus is also very light, but this might be a sign of a dynamically varying vacuum energy.


\section{Summary and Discussion}

We have discussed the stability of the volume of a warped codimension two
brane model with flux compactification,
by taking the one-loop quantum corrections of a massless scalar field into account.
We have also examined the {\it rigidity} of stability by comparing the ratio between comparison 
the effective modulus mass and the bulk curvature scales.

First, we introduced a warped flux compactification model in six dimensions, where we 
started from the solution with a spacetime structure similar to the Reisner-Nortstr${\rm \ddot{o}}$m black hole.
After a {\it double} Wick rotation, we were able to derive the desired
braneworld solution. 
The branes are located at the horizon positions and correspond to conical
singularities. For relatively low energies, namely when the brane tensions are less than the bulk gravitational energy scale, the brane tensions are related to the deficit angles directly. Once the brane tensions are fixed, via Eq. (\ref{tendef6d}), the ratio of the brane positions,
$r=\rho_-/\rho_+$ is also fixed.
But, the volume of the extra dimensions is not completely determined, because of the scale invariance of the solution. To this end, we investigated whether or not this dynamical degree of freedom can be fixed by considering the one-loop quantum effective action for a massless, minimally coupled bulk scalar field. We investigated the stability of the modulus for given values of the two brane tensions, or equivalently of $r$ and
$\delta_+$ (the deficit angle of the brane located at $\rho_+$).

The form of the effective potential can be fixed up to overall coefficients,
due to scale invariance and dimensional arguments, which depend on $r$ and $\delta_+$ as in Eq.~(\ref{originaleffpot}).
From this the condition for the volume modulus to have a minimum (and hence be stabilized) is determined only by the sign of the coefficient in the logarithmic term, i.e.,
$B_6(r,\delta_+)$. This coefficient is directly related to the zeta function 
$\zeta(0,\Delta_6)$ and thus, the heat kernel coefficient $a_6(f=1)$. 

In the actual investigation of stability, we have used an alternative
model in four dimensions that has a very similar spacetime structure to the original 
six-dimensional model. This is because the contribution for the cone  
to the $a_6$ heat kernel coefficient has {\it never} been formulated as far as
the authors are aware.
We then showed that the heat kernel coefficient is positive definite 
for most dilaton couplings, $g$, independently of the choice of $r$ and
$\delta_+$.  The contribution from the bulk is positive, whereas
that from the conical branes is negative for smaller $r$;
though they are of the same order.
For all of the cases with $g\sim {\cal O}(1)$, the bulk dominates over the brane 
parts and thus, the volume is {\it stabilized}.

However, one might ask whether or not the stability is {\it rigid} against the various KK perturbations. This rigidity can be determined from the ratio between the effective modulus mass and the stabilized curvature radius of the bulk, as given by Eq.~(\ref{massratio}), implying a knowledge of the ratio of $A_6(r,\delta_+)/B_6(r,\delta_+)$. If this ratio is less than unity,
then the mass of the volume modulus
becomes too light and may be destabilized by other KK contributions,
whereas if it is greater than
unity, the modulus field is not easily perturbed from the ``stabilized" minimum. 
We showed that the mass ratio becomes larger for larger coupling $g$, which
is proportional to $g^2$ and therefore, stability is {\it rigid}.
For larger degrees of warping, i.e., for smaller value of $r$, other than $r=0$,
we found that the stabilized modulus mass becomes relatively light and that this ratio is not so sensitive to different values of deficit angle $\delta_+$. Note that it only exhibits a divergence at $\delta_+=2\pi$, where the bulk essentially vanishes.

Then, we made some suggestions for the six-dimensional case and by comparing the bulk part with that of the four-dimensional $a_4$ bulk part we found similar behavior indicating that our conclusion in four dimensions will carry through to the six-dimensional case: The modulus can be stabilized and furthermore our approximate WKB approach suggests that the mass of the modulus leads to rigidity for certain couplings and parameters.  Of course, once formal expressions for the six-dimensional conical heat kernel coefficients are found, then we can extend the same analysis as that employed in this paper. However, a more accurate (semi-analytical/numerical) method to evaluate $\zeta'(0, \Delta_6)$ should be found, rather than the approximate WKB eigenvalue method that we used in this article. 

We also discussed some phenomenological issues, i.e., the hierarchy problem of the 
fundamental energy scales and the extremely tiny value of the observed vacuum
energy density. The six-dimensional results were extrapolated from the 
corresponding results in four dimensions.
We found that in the hierarchy problem, the ratio between the energy scale
of the brane localized field and the gravitational scale becomes
much smaller than unity, especially for larger degrees of warping $r\ll 1$.
Furthermore, the effective vacuum energy density can also be much smaller
than the gravitational scale. The vacuum energy is negative by definition and we therefore 
need some kind of uplifting mechanism, to obtain a more realistic cosmology. 
In these cases, as was mentioned previously, the mass of the volume modulus may also be small 
and thus, might be destabilized.

It is also interesting to compare our results with the codimension one case.  
An important difference is that in the codimension one case all the moduli are
not dynamically fixed
 and should be stabilized by quantum corrections of bulk fields;
however in the codimension two case at least one of the moduli fields 
is fixed dynamically, because the brane tensions are directly related to 
the bulk geometry. 
In terms of heat kernel coefficients another obvious difference is that we are considering even 
dimensions. In odd dimensions only boundary terms occur and thus,
the renormalization only affects the brane tensions, whereas in even dimensions there is a bulk contribution
and hence logarithmic terms appear. 
Probably the most important difference is that the stabilization of the volume modulus
can be realized by the contribution from the bulk, not from the branes,
for small warpings.
For $r\ll 1$ (large warpings) the stabilized modulus effectively behaves as a very light scalar field on the brane, which is somewhat similar to the codimension one case, e.g., see ~\cite{Garriga:2000jb,Garriga:2001ar}. 

Before closing this article we shall refer to some extensions of this work:
 One issue is to do with the quantum corrections of other fields, such as the magnetic flux and  graviton. The tensor part of the graviton perturbations should be equivalent to a minimally coupled scalar field, which we investigated in this article and 
such analyses might give us important insights into warped flux compactifications with self-gravitating branes. Besides quantum effects, there are many interesting issues that still remain in codimension two braneworld, e.g., gravitational waves in the bulk~\cite{deRham:2005ci}, 
non-linear dynamics of the bulk and branes, thick branes and so on.
Such work might be key to understanding the self-tuning mechanism of the effective cosmological constant as a dynamical process.
We hope to report on these topics in ongoing work \cite{mns, mm}.

\section*{Acknowledgements}
We are grateful to A. Flachi and K. Uzawa for valuable and interesting discussions.
This work was supported in part by Monbukagakusho Grant-in-Aid for Scientific Research(S) No. 14102004 and (B) No.~17340075.
M.~M. and M.~S. were also supported by the Japan-France Research Cooperative Program.

\appendix
\section{A four-dimensional counterpart}

\subsection{The model}

In this Appendix, we consider an alternative model of a warped
flux compactification in a four-dimensional Einstein-Maxwell-dilaton theory
given by Eq. (\ref{theory}).
 This is a toy-model analogue of the original six-dimensional co-dimension two braneworld model with a warped flux compactification  and a non-vanishing scalar potential \cite{Burgess:2003mk}:
Hereafter, we set the bulk gravitational scale to $M_4^2=1$ for simplicity;
we shall reinsert it if and when it is needed.


As for six dimensions, we first consider a black hole solution of the theory \cite{Aghababaie:2003ar}. After a {\it double} Wick rotation,
 we obtain
\begin{eqnarray}
ds^2&= & h(\rho)d\theta^2
     +\frac{d\rho^2}{h(\rho)}
     +(2\rho)(-d\tau^2+d x_2^2)\,,
     \nonumber\\
     &&h(\rho)=2 M-\frac{Q^2}{\rho}-2g^2\rho\,,
     \nonumber\\
     &&\varphi(\rho)=-\ln(2\rho)\,,
     \nonumber\\
     &&F_{\theta\rho}=- \frac{ Q}{\rho^2}
\,.
\end{eqnarray}
In this scenario the branes are located at positions that are determined by
the {\it horizon} condition $h(\rho)=0$;
\begin{eqnarray}
 \rho_{\pm}
=\frac{M}{2g^2}
  \left(1\pm \sqrt{1-\frac{2 g^2 Q^2}{M^2}} \right)\,.
\end{eqnarray}
We can then rewrite
\begin{eqnarray}
h(\rho)=\frac{2g^2}{\rho}
       \left(\rho_+ -\rho\right)
       \left(\rho -\rho_-\right)
\end{eqnarray}
and whence we also obtain the following useful relations
\begin{eqnarray}
\rho_+ + \rho_- = \frac{M}{g^2}\,,
\qquad
\rho_+\rho_-= \frac{Q^2}{2g^2}\,.
\end{eqnarray}
This gives a direct relation between the mass and charge to the magnetic 
flux. The global period of $\theta$, denoted $\Delta \theta$, is
determined by the brane tensions through the tension-deficit relations as is 
discussed later.

After a coordinate transformation 
\begin{eqnarray}
z=\left(\frac{\rho_+ - \rho}{\rho - \rho_-}\right)^{1/2}
\end{eqnarray}
the bulk metric becomes
\begin{eqnarray}
ds^2= 2\rho\,
\left(\frac{dz^2}{g^2(1+z^2)^2}
     +g^2\frac{(\rho_+-\rho_-)^2 z^2}{(\rho_++\rho_-z^2)^2}d\theta^2
         -d\tau^2 + dx_2^2
\right)\,,
\label{bkgrdzmetric}
\end{eqnarray}
where our choice of factorization is for later convenience. In this frame we can now easily investigate the asymptotic behavior of the spacetime near 
each of the conical branes.
For $z\to 0, \infty$, there are cones whose deficit angles are obtained from
\begin{eqnarray}
\delta_{+}:= 2\pi-g^2 (1-r)\Delta\theta\,,
\qquad\qquad
\delta_{-}:= 
2\pi-g^2 \frac{1-r}{r}\Delta \theta
 \,.\label{conic3}
\end{eqnarray}

The modulus corresponds to the proper length between two branes (strings), which is characterized by $\rho_{\pm}$, or equivalently
$\rho_+$ and $r$. The string tensions are related to the conical deficits by
\begin{eqnarray}
\sigma_{\pm}=  M_4^2 \delta_{\pm}\,.
\end{eqnarray}
We stress that these relations are only valid for sufficiently small brane tensions in comparison with the bulk scale $M_{4}^2$.
Again, $\Delta\theta$ is given by a relation identical to Eq. (\ref{angle period}) and thus, 
\begin{eqnarray}
 \frac{2\pi-\delta_+}{2\pi-\delta_-}=r\,.\label{tendef4d}
\end{eqnarray}
After eliminating $r$, we obtain
\begin{eqnarray}
 \Delta\theta
=\frac{(2\pi-\delta_+)(2\pi-\delta_-)}
      {g^2(\delta_+-\delta_-)}
=\frac{2\pi-\delta_+}{g^2(1-r)}
      \,.
\end{eqnarray}

Once the brane tensions, $\sigma_+$ and $\sigma_-$ are fixed,
then $r$ is also fixed and so we now regard the free parameters as 
$r$ and $\sigma_+$, along with the dilaton bulk coupling $g$.
The remaining degree of freedom used to determine the bulk geometry is 
the absolute size of the bulk, i.e., $\rho_+$. 
However, because of the scale invariance, $\rho_+$
can only be fixed by quantum corrections of bulk fields.
The magnetic flux is given by
\begin{eqnarray}
\int^{\rho_+}_{\rho_-}d\rho
\int^{\Delta \theta}_0 d\theta
F_{\theta\rho}
=-\Delta\theta Q
\Big(\frac{1}{\rho_-}-\frac{1}{\rho_+}\Big)
=-\frac{2\pi-\delta_+}{g}\sqrt{\frac{2}{r}}\,.
\end{eqnarray}
Thus, the gauge field flux does not contribute to the size modulus.

\subsection{Massless scalar field perturbations in a four-dimensional model}

We discuss the basic properties of a massless, minimally coupled scalar field
on this background, which is introduced
in order to investigate the one-loop quantum effects in the counterpart model discussed 
in Sec IV and V, whose action is given by
\begin{eqnarray}
S_{\rm scalar}=-\frac{1}{2}
                \int d^4 x\sqrt{g}
                \phi\Delta_4 \phi\,. 
    \label{scalaraction4d}
\end{eqnarray}
As before, we consider a continuous conformal transformation of the metric, parameterized by $\epsilon$;
\begin{eqnarray}
d{\tilde s}_{4,\epsilon}^2
 = e^{2(\epsilon-1)\omega}ds_4^2\,,
 \qquad
\omega= \frac{1}{2} \ln(2\rho)\,,\label{confort4d}
\end{eqnarray}
and thus
\begin{eqnarray}
 d{\tilde s}_4^2
=(2\rho)^{\epsilon}
 \Big(
   \frac{dz^2}{g^2(1+z^2)^2}
 +\frac{g^2(\rho_+ - \rho_-)^2 z^2}
       {(\rho_+ + \rho_- z^2)^2}
    d\theta^2
 + d{\bf x}^2
 \Big)\,,\label{conf4d}
\end{eqnarray}
where for $\epsilon=1$ we have the original metric, which we shall define as
$\Delta_{4,\epsilon}=\Delta_4$.
The classical action changes as 
\begin{eqnarray}
  S_{\rm scalar}
   =-\frac{1}{2}\int d^4 x\sqrt{g}
     \phi\Delta_4\phi
   =-\frac{1}{2}\int d^4 x\sqrt{\tilde g}
     \tilde \phi\left(\tilde \Delta_4+E_4(\epsilon) \right)
     \tilde \phi\,,
     \label{class4d}
\end{eqnarray}
where
\begin{eqnarray}
E_4
(\epsilon)
&=& -(\epsilon-1)^2 \tilde{g}^{ab}
                 \nabla_{a} \omega
                  \nabla_{b}\omega
     +(\epsilon-1){\tilde \Delta}_4\ln\omega  
   \nonumber\\
 &=&
\left(\frac{1}{2\rho}\right)^{\epsilon}
\frac{g^2(1-\epsilon)(\rho_+ - \rho_-) 
   \big\{
     \rho_+ (2+(1-\epsilon)z^2)
+  \rho_- z^2(-1+\epsilon-2z^2)
     \big\} }
     {(\rho_+ + \rho_- z^2)^2}\,.
\end{eqnarray}


\section{Conformal invariance of heat kernel coefficients}

The classical action of a massless, minimally coupled scalar field is changed under this conformal transformation, see Eq. (\ref{class4d}).
In four dimensions, including conical branes, the $a_4$ heat kernel coefficient is given by
\begin{eqnarray}
 a_4(f=1)
 &:=&(4\pi)^{-2}
  \Bigg\{360^{-1}
   \int_M d^4 x \sqrt{g}
  \left\{
      \left(60E_{4}{}^{;k}{}_{;k}
            +60RE_{4}
            +180E_{4}^2
            +12R^{;k}{}_{;k}
            +5R^2 
            -2R_{ij}R^{ij}
            +2R_{ijkl}R^{ijkl}
       \right)
   \right\}
 \nonumber\\
   &+&\int d^2 x \sum_{A=\pm}
  \Big(\frac{\delta_A}{2\pi}\Big)
    \frac{2-\frac{\delta_A}{2\pi}}
         {1-\frac{\delta_A}{2\pi}}
  \sqrt{h_A}\, 
\Big[
  \frac{\pi}{3}
  \Big( E_4   
        + \frac{1}{6}R
        +\lambda_1\sum_a
         (\kappa^{(a)}{}^2
                 -2 \kappa^{(a)}_{ij}\kappa_{(a)}^{ij}
          )
  \Big)
 \nonumber\\
  &-&\frac{\pi}{180}
  \frac{ 2-\frac{\delta_A}{\pi}+\frac{\delta_A^2}{4\pi^2}}
     {(
     1-\frac{\delta_A}{2\pi}
       )^2}
  \Big(\sum_a R_{aa}-2\sum_{a,b} R_{abab}
   +\frac{1}{2}\sum_a \kappa^{(a)}{}^2
   +\lambda_2 \sum_a
   (           \kappa^{(a)}{}^2
             -2 \kappa^{(a)}_{ij}\kappa_{(a)}^{ij}) 
   \Big)
   \Big]
   \Bigg\}\,.
\label{a4conf}
\end{eqnarray}

\subsection{Contribution of the bulk}
The integrand of the bulk part of Eq.~(\ref{a4conf}) is
\begin{eqnarray}
&&\big(B_4(r,\delta_+)\big)_{\rm bulk}
=
\frac{g^2(2\pi-\delta_+)}{1440\pi^2}
\int^{\infty}_0 dz
\frac{z}
     {(1+z^2)(1+r z^2)}
\frac{\Psi_4(1,r,\epsilon, z)}{(1+r z^2)^4}
\,. \label{bulk}
\end{eqnarray}
where
\begin{eqnarray}
\Psi(\rho_+,\rho_-,\epsilon,z)
 &=&
-216\,\epsilon{z}^{6} \rho_+^{3}\rho_-
-40\, \rho_+^{3}{\epsilon}^{2}{z}^{6}\rho_-
+504\,{z}^{6}\epsilon \rho_+^{2}\rho_-^{2}
-144\,{z}^{8}\epsilon\rho_+\,\rho_-^{3}
+48\,{z}^{8}\epsilon \rho_+^{2}\rho_-^{2}
+8\,{z}^{8}{\epsilon}^{2}\rho_+^{2} \rho_-^{2}
\nonumber\\
&-&48\, \rho_+^{2}{\epsilon}^{3}{z}^{6}\rho_-^{2}
-112\,\rho_+^{3}{\epsilon}^{2}{z}^{4}\rho_-
+192\,\rho_-^{2}\rho_+^{2}{\epsilon}^{2}{z}^{4}
-96\,\rho_+^{2}{\epsilon}^{3}{z}^{4} \rho_-^{2}
-88\,{\epsilon}^{2}{z}^{6}\rho_-^{3} \rho_+
\nonumber\\
&+&64\,\rho_+\,{\epsilon}^{3}{z}^{4}\rho_-^{3}
+33\,\rho_+^{4}{z}^{4}
+8\,{\epsilon}^{2} \rho_+^{4}
+33\, \rho_-^{4}{z}^{4}
+48\,\epsilon\rho_+^{2}\rho_-^{2}
-36\,\rho_+^{2}{z}^{2}\rho_-^{2}
+156\, \rho_+\,{z}^{4}\rho_-^{3}
\nonumber\\
&+&72\,\epsilon{z}^{4}\rho_-^{4}
-144\,\epsilon\rho_+^{3}\rho_-
+24\,{\epsilon}^{2}{z}^{2}\rho_+^{4}
-16\,\rho_+^{4}{\epsilon}^{3}{z}^{4}
+16\,{\epsilon}^{3}{z}^{2}\rho_-^{3} \rho_+
-48\,{\epsilon}^{3} \rho_+^{2}{z}^{2} \rho_-^{2}
\nonumber\\
&+&48\,{\epsilon}^{3}\rho_+^{3}{z}^{2}\rho_-
+104\,{z}^{6}{\epsilon}^{2} \rho_+^{2}\rho_-^{2}
-16\,{z}^{8}{\epsilon}^{2}\rho_+\, \rho_-^{3}
-16\,{\epsilon}^{3}{z}^{6}\rho_-^{4}
-16\,{\epsilon}^{3}{z}^{4}\rho_-^{4}
-16\,\rho_-\,{\epsilon}^{2} \rho_+^{3}
\nonumber\\
&+&168\,\epsilon{z}^{6}\rho_-^{4}
+8\,{\epsilon}^{2}{z}^{8}\rho_-^{4}
+8\,{\epsilon}^{2} \rho_-^{2}\rho_+^{2}
+96\,\epsilon{z}^{8}\rho_-^{4}
-16\,{\epsilon}^{3}\rho_+^{4}{z}^{2}
+16\, \rho_+^{4}{\epsilon}^{2}{z}^{4}
+60\,\rho_+^{3}{z}^{6}\rho_-
\nonumber\\
&+&24\,{\epsilon}^{2}{z}^{6} \rho_-^{4}
-88\,{\epsilon}^{2}{z}^{2}\rho_+^{3}\rho_-
+96\,\epsilon  \rho_+^{4}
+104\,{\epsilon}^{2}{z}^{2}\rho_+^{2}\rho_-^{2}
+84\,\rho_+^{4}{z}^{2}
-40\,{\epsilon}^{2}{z}^{2}\rho_+\,\rho_-^{3}
+276\,\rho_+^{3}{z}^{2}\rho_-
\nonumber\\
&+&198\, \rho_+^{2}{z}^{4}\rho_-^{2}
+156\, \rho_+^{3}{z}^{4} \rho_-
+504\,\epsilon{z}^{2}\rho_+^{2}\rho_-^{2}
-528\,\epsilon{z}^{4} \rho_+^{3} \rho_-
+912\,\epsilon{z}^{4}\rho_+^{2} \rho_-^{2}
-456\,\epsilon{z}^{2}\rho_+^{3} \rho_-
\nonumber\\
&+&16\,{\epsilon}^{2}{z}^{4}\rho_-^{4}
+72\,\epsilon{z}^{4}\rho_+^{4}
+16\,\rho_+^{3}{\epsilon}^{3}{z}^{6} \rho_-
+84\,\rho_-^{4}{z}^{6}
+64\,\rho_+^{3}{\epsilon}^{3}{z}^{4}\rho_-
+96\, \rho_-^{4}{z}^{8}
-216\,\epsilon{z}^{2} \rho_-^{3} \rho_+
\nonumber\\
&+&60\,\rho_-^{3}{z}^{2} \rho_+
-528\,\epsilon{z}^{4} \rho_+\,\rho_-^{3}
-456\,\epsilon{z}^{6} \rho_-^{3}\rho_+
+48\,\rho_+\,{\epsilon}^{3}{z}^{6} \rho_-^{3}
-36\,\rho_+^{2}{z}^{6} \rho_-^{2}
\nonumber\\
&-&112\,\rho_+\,{\epsilon}^{2}{z}^{4}\rho_-^{3}
+96\, \rho_+^{4}
+276\, \rho_-^{3}{z}^{6}\rho_+
+168\,\epsilon{z}^{2}\rho_+^{4}
\,.\label{psiconf}
\end{eqnarray}

At a glance, Eq. (\ref{bulk}) appears to depend on the parameter $\epsilon$;  however, we can show that it is independent of $\epsilon$, i.e., it is conformally invariant.

\subsection{Contribution of the conical branes}

We can also evaluate the part coming from the cones. Starting with the normal derivatives
\begin{eqnarray}
  n^{i}_{(z)}
= \left(\frac{1}{2\rho}\right)^{\epsilon/2}
  g(1+z^2)\delta^i_{z}
 \,,\qquad\qquad
   n^{i}_{(\theta)}
= \left(\frac{1}{2\rho}\right)^{\epsilon/2}
 \frac{\rho_+ + z^2\rho_-}
      {g (\rho_+ - \rho_-)z}
      \delta^i_{\theta}
\end{eqnarray}
in the bulk we obtain
\begin{eqnarray}
    \sum_a R_{aa}
- 2 \sum_{a,b} R_{abab}
&=&
-g^{2}
\frac{1}{(2\rho)^{\epsilon}}
\frac{1}{
\left (\rho_+ + \rho_-\,z^{2}\right )^{2}}
\nonumber
\\
&&\Big(
-4 z^{4}\rho_-^{2}
+12\,{\rho_+}\,{\rho_-}\,{z}^{4}
+2\rho_+^{2}\epsilon^{2}z^{2}
+2\epsilon^{2}z^{2} \rho_-^{2}
+24\,\rho_+\,\rho_-\,{z}^{2}
-4\,\rho_+^{2}z^{2}
\nonumber\\
&&-4\,z^{2}\rho_-^{2}
+12\,\rho_+\,\rho_-
-4\,\rho_+\,\epsilon^{2}z^{2}\rho_-\,
-4\,\rho_+^{2}\Big)\,,
\nonumber\\
E+\frac{1}{6}R
&=&
\frac{g^2}{3}
\frac{1}{(2\rho)^{\epsilon}}
\frac{1}{(\rho_+ +\rho_-\,z^{2})^{2}}
\Big (
4\,{{\rho_+}}^{2}
+6\,\rho_+\,\rho_-\,z^{2}
+\rho_+^{2}z^{2}
+{z}^{2}\rho_-^{2}
+4z^{4}\rho_-^{2}
\Big )\,.
\end{eqnarray}
These values are invariant for the limits, $z\to 0,\, \infty$. The extrinsic curvature in each direction is given by
\begin{eqnarray}
 \kappa_{ij}^{(z)}
=-\frac{1}{2}\left(2\rho\right)^{\epsilon/2}
  g(1+z^2) \partial_z
     \ln(2\rho)^{\epsilon}\delta_{ij}\,,
     \qquad\qquad
 \kappa_{ij}^{(\theta)}=0\,
\end{eqnarray}
and we find
\begin{eqnarray}
\sum_a \kappa^{(a)}{}^2
= \kappa^{(z)}{}^2
 +\kappa^{(\theta)}{}^2
&=&16(2\rho)^{-\epsilon}
  \epsilon^2 g^2
 \frac{ \left(\rho_+-\rho_-\right)^2 z^2}
      {\left(\rho_+ + \rho_- z^2\right)^2 }\,,
\nonumber\\
\sum_a
\Big(
 \kappa^{(a)}{}^2
 -2 \kappa^{(a)}_{ij}\kappa_{(a)}^{ij}
 \Big)
&=& 8 \left(2\rho\right)^{-\epsilon}
\epsilon^2 g^2
\frac{(\rho_+-\rho_-)^2 z^2}{(\rho_+ +\rho_- z^2)^2}\,.
\label{extzero}
\end{eqnarray}
Thus, we see that these terms vanish at the locations of the 
conical branes: at $z=0$ and $z=\infty$.

Given this fact it is easy to show that $a_4(f=1)$ is also conformally invariant.
The explicit form of $B_4(r,\delta_+)$ for conical branes is finally obtained as
\begin{eqnarray}
 \big(B_4(r,\delta_+)\big)_{\rm branes}
&=& \frac{g^2}{288\pi}
\frac{\delta_+}{2\pi}
\frac{2-\frac{\delta_+}{2\pi}}
     {1-\frac{\delta_+}{2\pi}}
 \Big[4
      -\frac{1}{5}
       \frac{1+(1-\frac{\delta_+}{2\pi})^2}
            {(1-\frac{\delta_+}{2\pi})^2}
          (1-3r)
 \Big]
 \nonumber\\
&&
+  \frac{g^2}{288\pi}
\frac{\frac{\delta_+}{2\pi}-(1-r)}{r}
\frac{r+(1-\frac{\delta_+}{2\pi})}
     {1-\frac{\delta_+}{2\pi}}
 \Big[4
      -\frac{1}{5}
       \frac{r^2+(1-\frac{\delta_+}{2\pi})^2}
            {(1-\frac{\delta_+}{2\pi})^2}
          (1-\frac{3}{r})
 \Big]\,,\label{inflationary}
\end{eqnarray}
where we have used
\begin{eqnarray}
\delta_- =\frac{\delta_+ - 2\pi(1-r)}{r}\,.
\end{eqnarray}

We have also confirmed that the bulk part of $a_6(f=1)$ is conformally
invariant and we expect a similar behavior for the cone part, as in four dimensions.

\section{Derivation of cocycle function for conical boundaries}

 In this Appendix, we shall give a derivation of the cocycle function 
 for conical boundaries in four dimensions. In deriving such a term, we shall 
use the conformal properties of  the heat kernel coefficients, see e.g., \cite{Dowker:1989ue,Dowker:1994bj}~.

 \subsection{Conformal properties of the heat kernel coefficients}

In general, for a conformal transformation of the metric, given by
$g_{ij} \to e^{2\omega}g_{ij}$, the heat kernel coefficient in 
$d$-dimensions satisfies the following relation
\begin{eqnarray}
 a^{(d)}_{k} [g_{ij}, \delta \omega]
=\frac{1}{d-2k}
 \delta a^{(d)}_{k} 
    [e^{2\omega} g_{ij},1]\Big|_{\omega=0}\,. \label{heatconf}
\end{eqnarray}
We shall use this formal relation to derive the {\it smeared} heat kernel coefficients.
At a glance, there seems to be a pole at $d=2k$ in Eq.~(\ref{heatconf}). 
However, because $ a^{(d)}_{k}(f=1)$ is conformally invariant in four-dimensions 
(see Appendix B), then the leading order pole term vanishes.
If $a^{(d)}_{k}(f=1) $ is {\it not} conformally invariant, then we should add the term, $\,\delta a^{(d)}_{k-2}[\delta \omega]|_{\omega=0}$, which includes mass and curvature couplings \cite{Dowker:1989ue}.
This is in order to keep conformal invariance on the right-hand-side.
In our case $a^{(d)}_{k} (f=1)$ is already conformally invariant, because of the $E$ term that appears as a result of the conformal transformation of the classical scalar action.

The conical contribution to $a_4(f=1)$ is given by
\begin{eqnarray}
\label{a4f1}
 a_{4,{\rm cone}} (f=1)
 &:=&(4\pi)^{-2}
  \int d^2 x \sum_{A=\pm}
  \left(\frac{\delta_A}{2\pi}\right) 
  \frac{2-\frac{\delta_A}{2\pi}}
       {1-\frac{\delta_A}{2\pi}}
  \sqrt{h_A}\,
\Big[
  \frac{\pi}{3}
  \Big( E_4   
        + \frac{1}{6}R
        +\lambda_1 \sum_a
         \big(\kappa^{(a)}{}^2
           -2 \kappa^{(a)}_{ij}\kappa_{(a)}^{ij}
          \big)
  \Big)
 \nonumber\\
  &-&\frac{\pi}{180}
  \frac{(2-\frac{\delta_A}{\pi}+\frac{\delta_A^2}{4\pi^2})}
     {(1-\frac{\delta_A}{2\pi})^2}
 \Big(\sum_a R_{aa}-2\sum_{a,b} R_{abab}
   +\frac{1}{2}\sum_a \kappa^{(a)}{}^2
   +\lambda_2 \sum_a
   \big( \kappa^{(a)}{}^2
             -2 \kappa^{(a)}_{ij}\kappa_{(a)}^{ij}
   \big)
   \Big)
 \Big]
\end{eqnarray}
and in our case the conformal transformation is defined by 
Eq. (\ref{confort4d}).

For a general $d$-dimensional spacetime, we obtain
\begin{eqnarray}
E_d(\epsilon) 
 &=&\Big(\frac{d-2}{2}\Big)
  (\epsilon-1){\tilde\Delta} \omega
  -\Big(\frac{d-2}{2}\Big)^2
  (\epsilon-1)^2
   {\tilde\nabla}^i \omega
   {\tilde\nabla}_i \omega\,.
\nonumber\\
&=&e^{-2(\epsilon-1)\omega}
  \Big[
  \Big(\frac{d-2}{2}\Big)
  (\epsilon-1){\Delta} \omega
  +\Big(\frac{d-2}{2}\Big)^2
  (\epsilon-1)^2
   {\nabla}^i \omega
   {\nabla}_i \omega
  \Big]
\,,
\end{eqnarray}
where in the second step we have used the conformal scaling property of the 
Laplacian, e.g., see \cite{kirs}. Note that $\epsilon=1$ corresponds to the original metric for the codimension two brane. Given the above conformal transformation the Ricci scalar, Ricci and Riemann curvature terms
scale as \cite{kirs, Dowker:1994bj}
\begin{eqnarray}
\Big( E_d+\frac{1}{6}R \Big) (\epsilon)
&=& e^{-2(\epsilon-1)\omega}
 \Big[\frac{1}{6} R
+\frac{1}{6}(\epsilon-1)(d-4) \Delta \omega
+\frac{1}{12}(\epsilon-1)^2(d-2)(d-4) \omega_{;k}\omega^{;k}
 \Big]
 \nonumber\\
 \Big(\sum_a R_{aa}
     -2\sum_{a,b} R_{abab}
 +\frac{1}{2}\sum_a \kappa^{(a)}
             \kappa^{(a)}\Big)(\epsilon)
 &=&e^{-2(\epsilon-1)\omega}
 \Big[
  \sum_a  R_{aa}
  -2\sum_{a,b} R_{abab}
  +\frac{1}{2}
     \sum_{a,b} \kappa^{(a)}\kappa^{(a)}
\nonumber\\
 &&-(\epsilon-1)
  \Big( 
  (d-4) \sum_a  \omega_{;ij}n^{(a)i}n^{(a)j}
  +2 \Delta_2 \omega
  +(d-2)\sum_a \kappa^{(a)}\omega_{;i} n^{(a)i}
  \Big)
\nonumber\\
&&+(\epsilon-1)^2
 \Big(
 \frac{1}{2}(d-4)(d+2) 
   \sum_a \omega_{;i}\omega_{;j}n^{(a)i} n^{(a)j}
    -2(d-4)\omega_{;k}\omega^{;k}
 \Big)
 \Big]
 \nonumber\\
\sum_a 
\Big( \kappa^{(a)}{}^2
  -2 \kappa^{(a)}_{ij}\kappa_{(a)}^{ij}
\big)
\Big)(\epsilon)
 &=&
 e^{-2(\epsilon-1)\omega} \sum_a
  \Big[
    \kappa^{(a)}{}^2
     -2 \kappa^{(a)}_{ij}\kappa_{(a)}^{ij}
 -2(d-4)(\epsilon - 1)\omega_{;i}n^{(a) i} \kappa^{(a)} 
\nonumber\\
&&+(d-2)(d-4)(\epsilon-1)^2 
       \omega_{;i} \omega_{;j} n^{(a)i} n^{(a)j}\Big]\,,
       \label{coldspot}
\end{eqnarray} 
where $;$ denotes the covariant derivative with respect to the metric
$\epsilon=1$ and curvature tensors which appear on the right-hand-side
 are also defined for the same metric.
Here, $\Delta_2$ represents the Laplace operator defined by
\begin{eqnarray}
\Delta_2 \omega = \Delta \omega 
                  -\sum_a n^{(a)i}n^{(a)j} \omega_{;ij}\,. 
\end{eqnarray}
Note that terms including $\Delta_2 \omega$ and $\kappa^{(a)}\omega_{;i} n^{(a)i}$ vanish on the cone in the model which we are considering and thus, we neglect these terms. 
All the remaining terms proportional to $(\epsilon-1)$ are
also proportional to $(d-4)$. 
The terms which include $\kappa^{(a)i}{}_{j}$ (or $\kappa^{(a)}$) are also
vanishing on the cones. However, in order to see the conformal properties of these terms,
in this subsection, we keep them.

By multiplying the conformal transformation of the trace of the brane induced metric,  $\sqrt{h(\epsilon)}:=e^{(d-2)(\epsilon-1)\omega} \sqrt{h}$, with these combinations of intrinsic and extrinsic curvature terms, we can expand the terms as a series in powers of ($(d-4)$). 
For instance, expanding the first term in Eq. (\ref{a4f1}) we obtain 
\begin{eqnarray}
&&\sqrt{h(\epsilon)}
\Big[
 E_d + \frac{1}{6}R
 +\lambda_1\sum_a
   \Big(\kappa^{(a)}{}^2
       -2 \kappa^{(a)}_{ij}\kappa_{(a)}^{ij}\Big)
\Big]
\nonumber\\
&=&
 \sqrt{h}
 e^{(d-4)(\epsilon-1)\omega}
 \Big[
 \frac{1}{6} R
 +\lambda_{1}\sum_a
 \Big(\kappa^{(a)}{}^2
   -2 \kappa^{(a)}_{ij}\kappa_{(a)}^{ij}\Big)
\nonumber \\
&&
+(d-4)
\Big\{
\frac{1}{6}(\epsilon-1) \Delta \omega
+\frac{1}{12}(\epsilon-1)^2(d-2) 
       \omega_{;k}\omega^{;k}
+\lambda_1
\Big(
 -2(\epsilon-1)\sum_a
   \omega_{;i}n^{(a)i}\kappa^{(a)}
\nonumber\\
&& +(d-2)(\epsilon-1)^2\sum_a
    \omega_{;i}\omega_{;j}n^{(a)i}n^{(a)j}
\Big)
\Big\}
  \Big]
\nonumber\\
&=&
\sqrt{h}\Big[
 \frac{1}{6} R
 +\lambda_{1}\sum_a
 \Big( \kappa^{(a)}{}^2
   -2 \kappa^{(a)}_{ij}\kappa_{(a)}^{ij}\Big)\Big]
\nonumber\\
&+& (d-4)\sqrt{h}
 \Big[
   (\epsilon-1)\omega
   \Big(\frac{1}{6} R
 +\lambda_{1}\sum_a
 \Big(
 \kappa^{(a)}{}^2  
 -2 \kappa^{(a)}_{ij}\kappa_{(a)}^{ij}
 \Big)
 \Big)
 +\frac{1}{6}(\epsilon-1) \omega_{;k}{}^{;k}
 +\frac{1}{12}(d-2)(\epsilon-1)^2 \omega_{;k}\omega^{;k}
 \nonumber\\
 &&
 +\lambda_1
  \Big(
 -2(\epsilon - 1)\sum_a \omega_{;i}n^{(a) i} \kappa^{(a)} 
 +(d-2)(\epsilon-1)^2 
     \sum_a  \omega_{;i} \omega_{;j} n^{(a)i} n^{(a)j}  
  \Big)
 \Big]
 + O((d-4)^2)\,.
\end{eqnarray}   
Similarly, expanding the second piece in Eq. (\ref{a4f1}) leads to:
\begin{eqnarray}
&&
\sqrt{h}
\Big(
\sum_a R_{aa}
  -2\sum_{a,b} R_{abab}
  +\frac{1}{2}\sum_a \kappa^{(a)}{}^2
+\lambda_2 \sum_a
  \Big(
      \kappa^{(a)}{}^2
  -2 \kappa^{(a)}_{ij}\kappa_{(a)}^{ij}
 \Big)
\Big)
\nn
&+&
(d-4)
\sqrt{h}
\Big[
(\epsilon-1)\omega 
\Big(
\sum_a R_{aa}
  -2\sum_{a,b} R_{abab}
  +\frac{1}{2}\sum_a \kappa^{(a)}{}^2
+\lambda_2 \sum_a
  \Big(
      \kappa^{(a)}{}^2
  -2 \kappa^{(a)}_{ij}\kappa_{(a)}^{ij}
 \Big)
\Big)
\nonumber\\
&&-(\epsilon-1)\sum_a
    \omega_{;ij} n^{(a)i} n^{(a)j}
  +(\epsilon-1)^2
  \Big(
  \frac{1}{2}(d+2) 
   \sum_a \omega_{;i}\omega_{;j} n^{(a)i} n^{(a)j}
 -2 \omega_{;k} \omega^{;k}
  \Big)  
  \nonumber\\
 &&+\lambda_2
  \Big(
    -2(\epsilon-1)\sum_a \omega_{;i}n^{(a)i} \kappa^{(a)}
    +(d-2)(\epsilon-1)^2 \sum_a
       \omega_{;i}\omega_{;j} n^{(a) i} n^{(a)j}
   \Big)
\Big]
+O((d-4)^2)
\,.
\end{eqnarray}

\subsection{Cocycle function for the conical branes}

Now we can construct the cocycle function for the conical boundaries.
The $\sqrt{h} \omega_{;j}\omega^{;j}$ terms can be reduced 
to $\sum_{a}\omega_{;i}\omega_{;j}n^{(a)i}n^{(a)j}$ and 
because 
\begin{eqnarray}
  \sum_a
 \sqrt{h} \omega_{;i}  \omega_{;j}
          n^{(a)i}     n^{(a)j}
=\frac{z^2(\rho_+ - \rho_-)^2g^2}
      {(\rho_+ + \rho_- z^2)^2}
\end{eqnarray}
vanishes on the cone, at $z\to 0, \,\, \infty$, we can discard these terms. 
Whence
\begin{eqnarray}
a_{4,{\rm cone}} (g_{ij},
  \partial_{\epsilon}((\epsilon-1) \omega))
 =a_{4,{\rm cone}} (g_{ij},\omega)
&=&(4\pi)^{-2}
  \int d^2 x \sum_{A=\pm}
  \left(\frac{\delta_A}{2\pi}\right) 
   \frac{2-\frac{\delta_A}{2\pi}}
       {1-\frac{\delta_A}{2\pi}}
  \sqrt{h_A}\,
\Big\{
 \frac{\pi}{3}
   \Big(
   \frac{1}{6}\omega  R
  +\frac{1}{6}
    \omega_{;k}{}^{;k}
 \Big)
 \\
 &-&
 \frac{\pi}{180}
  \frac{ 
        \left(
     2-\frac{\delta_A}{\pi}
         +\frac{\delta_A^2}{4\pi^2}
   \right)}
     {\left(
     1-\frac{\delta_A}{2\pi}
   \right)^2}
\Big[
  \omega
 \big(\sum_a R_{aa}-2 \sum_{a,b} R_{abab}\big)
 -\sum_a \omega_{;ij}n^{(a)i}n^{(a)j} 
 \Big]
 \Big\}\,.
 \nonumber
\end{eqnarray}
Finally, performing the integration over $\epsilon$ we obtain
the cocycle contribution for the conical branes:
\begin{eqnarray}
&&-\int_0^1 d\epsilon\,
 a_{4,{\rm cone}} (g_{ij},\omega)
\nonumber\\
&=&-(4\pi)^{-2}
  \int d^2 x \sum_{A=\pm}
  \left(\frac{\delta_A}{2\pi}\right) 
  \sqrt{h_A}\,
\Big\{
 \frac{\pi}{3}
  \frac{2-\frac{\delta_A}{2\pi}}
       {1-\frac{\delta_A}{2\pi}}
\Big(\frac{1}{6}\omega R
  +\frac{1}{6}\sum_a \omega_{;ij}n^{(a)i}n^{(a)j}
  \Big)
\nonumber\\
&-&\frac{\pi}{180}
  \frac{ 
   ( 2-\frac{\delta_A}{2\pi})
        (
     2-\frac{\delta_A}{\pi}+\frac{\delta_A^2}{4\pi^2})}
     {(1-\frac{\delta_A}{2\pi})^3}
 \omega 
 \Big[
 \big(\sum_a R_{aa}-2\sum_{a,b} R_{abab}\big)
  -\sum_{a}\omega_{;ij}n^{(a)i}n^{(a)j} 
 \Big]
 \Big\}
\nonumber\\
&=&\int d^2 x 
\Big\{
-\frac{g^2(2\pi-\delta_+)}{1440\pi^2}
\int_0^1 d\epsilon
\int^{\infty}_0 dz
\Big(\frac{1}{2}\ln \Big(\frac{2\rho}{\rho_+}\Big) \Big)
\frac{ z}
     {(1+z^2)(1+r z^2)}
\frac{\Psi_4(1,r,\epsilon, z)}{(1+ r z^2)^4}
\nonumber\\
&-&\frac{g^2}{144\pi}
\frac{\delta_+}{2\pi}
\frac{2-\frac{\delta_+}{2\pi}}
     {1-\frac{\delta_+}{2\pi}}
 \Big[2 \ln(2\rho_+)
       -1+r
      -\frac{1}{5}
       \frac{1+(1-\frac{\delta_+}{2\pi})^2}
            {(1-\frac{\delta_+}{2\pi})^2}
        \Big( 
         \frac{1}{2}\ln(2\rho_+)  (1-3r) 
          +\frac{1-r}{2} 
        \Big)
 \Big]
 \nonumber\\
&-& \frac{g^2}{144\pi}
\frac{\frac{\delta_+}{2\pi}-(1-r)}{r}
\frac{r+(1-\frac{\delta_+}{2\pi})}
     {1-\frac{\delta_+}{2\pi}}
\nonumber\\
&\times&
 \Big[2\ln(2\rho_-)
       -1+\frac{1}{r}
      -\frac{1}{5}
       \frac{r^2+(1-\frac{\delta_+}{2\pi})^2}
            {(1-\frac{\delta_+}{2\pi})^2}
          \Big(
           \frac{1}{2}\ln(2\rho_-) (1-\frac{3}{r})
            +\frac{1-\frac{1}{r}}{2} 
          \Big)
 \Big]
 \Big\} \,.
\end{eqnarray}

\subsection{Total cocycle contribution}

Combining the above result with the bulk piece, we obtain the total cocycle part of
$A_4(r,\delta_+)$, defined in Eq. (\ref{4dcoef}), 
as
\begin{eqnarray}
 A_{4, {\rm cocycle}}(r,\delta_+)
&=&
-\frac{g^2(2\pi-\delta_+)}{1440\pi^2}
\int_0^1 d\epsilon
\int^{\infty}_0 dz
\Big(\frac{1}{2}\ln \Big(\frac{2\rho}{\rho_+}\Big) \Big)
\frac{z}
     {(1+z^2)(1+r z^2)}
\frac{\Psi_4(1,r,\epsilon, z)}{(1+r z^2)^4}
\nonumber\\
&-&\frac{g^2}{144\pi}
\frac{\delta_+}{2\pi}
\frac{2-\frac{\delta_+}{2\pi}}
     {1-\frac{\delta_+}{2\pi}}
 \Big[2 \ln(2)
       -1+r 
      -\frac{1}{5}
       \frac{1+(1-\frac{\delta_+}{2\pi})^2}
            {(1-\frac{\delta_+}{2\pi})^2}
        \Big( 
         \frac{1}{2}\ln (2)  (1-3r) 
          +\frac{1-r}{2} 
        \Big)
 \Big]
 \nonumber\\
&-& \frac{g^2}{144\pi}
\frac{\frac{\delta_+}{2\pi}-(1-r)}{r}
\frac{r+(1-\frac{\delta_+}{2\pi})}
     {1-\frac{\delta_+}{2\pi}}
\nonumber\\
&\times&
 \Big[2\ln(2r)
       -1+\frac{1}{r}
      -\frac{1}{5}
       \frac{r^2+(1-\frac{\delta_+}{2\pi})^2}
            {(1-\frac{\delta_+}{2\pi})^2}
          \Big(
           \frac{1}{2}\ln(2r) (1-\frac{3}{r})
            +\frac{1-\frac{1}{r}}{2} 
          \Big)
 \Big] \,,  \label{inflationary2}
\end{eqnarray}
where $\Psi_4(1,r,\epsilon, z)$ is given in Eq. (\ref{psiconf}).

\section{WKB analysis in the conformal frame}

\subsection{WKB mass spectrum}

In this Appendix, we derive the mass spectrum in 
the conformal frame ($\epsilon=0$): 
\begin{eqnarray}
 ds_4^2
=\frac{dz^2}{g^2(1+z^2)^2}
+\frac{g^2(\rho_+ - \rho_-)^2 z^2}
      {(\rho_+ + \rho_- z^2)^2}d\theta^2
+d{\bf x}^2 
\end{eqnarray}
and we consider the corresponding eigenvalue problem
\begin{eqnarray}
 \left(
  \Delta_4 
+ E(0)
  \right)
  \phi_{\lambda} =
-\lambda^2\phi_{\lambda}\,,
\end{eqnarray}
where $E(0)$ is obtained from Eq.~(\ref{good}) and we shall 
decompose the mass eigenfunction as
\begin{eqnarray}
 \phi_{\lambda}
=\int 
  \frac{d^2k}{(2\pi)}
  \sum_{m,n}
  \Phi_{\lambda}(z)e^{in(2\pi/\Delta\theta)\theta} e^{i {\bf kx }}\,.
\end{eqnarray}
From now on, we concentrate on the radial mode function and 
derive the mass spectrum in the context of the WKB approximation.
Changing variables to
\begin{eqnarray}
\Phi(z)= \left(\frac{\rho_+ +\rho_- z^2}
                 {z(1+z^2)}\right)^{1/2}
                 f(z)\,,
\end{eqnarray}
leads to the diagonalized equation of motion in terms of $f$:
\begin{eqnarray}
f''(z)+ q(z) f(z)=0\,,\label{bulkquintom}
\end{eqnarray}
where 
\begin{eqnarray}
 q(z)
=\frac{\lambda^2-k^2}{g^2(1+z^2)^2}
 -\frac{(1+ r z^2)^2 \alpha^2 n^2}
       {(1+z^2)^2 z^2 }
 +\frac{1}{4z^2}\,,\qquad\qquad
 \alpha:= \frac{2\pi}{2\pi-\delta_+}\,.
\end{eqnarray}
To apply the WKB method we require a Langer transformation \cite{langer} $y=\ln z$ 
and a further redefinition of the eigenfunction $F= e^{-y/2} f$.
Thus, Eq.~(\ref{bulkquintom}) becomes
\begin{eqnarray}
\left(
   \partial_y^2 
   +Q(y)
\right)F(y)=0\,,
\end{eqnarray}
where 
\begin{eqnarray}
   Q(y)
:= \frac{\lambda^2-k^2}{g^2(e^y+ e^{-y})^2}
 -\frac{( e^{-y}+ r e^{y})^2 \alpha^2 n^2}
       {(e^y+e^{-y})^2}\,.
\end{eqnarray}
The coordinate $y$ runs from $-\infty$ to $\infty$, while the turning points, defined by $Q(y_{\pm})=0$, are given by 
\begin{eqnarray}
  e^{2y_{\pm}}
=\frac{\frac{\lambda^2-k^2}{\alpha^2 g^2 n^2 }
       -2r
        \mp\sqrt{
        \Big(\frac{\lambda^2-k^2}{\alpha^2 g^2 n^2 }
       - 2r\Big)^2
       -4r^2}
         }
      {2r^2}\,,
\end{eqnarray}
where $y_{\pm}$ satisfies $y_-> 0 >y_+$. (In the original frame, $\epsilon=1$, the turning point problem is even more difficult to solve.) 
Furthermore, for the $n=0$ modes the turning points extend to $y\to \pm \infty$ and in 
such a case the mass spectrum becomes exact in the context of the WKB method. 
For $n\neq 0$, turning points appear if and only if 
\begin{eqnarray}
\lambda^2 >  k^2 
          + 4r \alpha^2 g^2 n^2\,,   \label{conde}
\end{eqnarray}
The mass spectrum, which will be determined later via the WKB method, is certainly valid for the modes that satisfy the above condition. However, we will use the WKB result for deriving the full zeta function, because this is the best we can do at present. Nevertheless, it should give a useful order of magnitude approximation to $\zeta'(0,\Delta_{4,\epsilon=0})$.

In the WKB approximation, the mass spectrum is given by the quantization
condition
\begin{eqnarray}
 \int_{y_+}^{y_-}dy
      \sqrt{Q(y)}
= \big(m+\frac{1}{2}\big)\pi\,,
\end{eqnarray}
where the integration can be performed as follows: 
First, we change the integration variable
\begin{eqnarray}
 \int dy \sqrt{Q(y)}
=\int \frac{du}{2u}
  \frac{\sqrt{\frac{(\lambda^2-k^2)}{g^2}u 
        -n^2 \alpha^2(1+ ru)^2 }}
       {1+u}\,,
\label{wkbint}
\end{eqnarray}
where $u=e^{2y}$. Then, we use the fact that
\begin{eqnarray}
 \int du \,\frac{\sqrt{a+bu+cu^2}}{u(1+u)}
&=&
\sqrt{-c}
 \arcsin\left(\frac{2cu+b}{\sqrt{-\Delta}}\right)
-\sqrt{-a}
 \arcsin\left(\frac{2au^{-1}+b}{\sqrt{-\Delta}}\right)
\nonumber\\
&-&(a-b+c)\int \frac{du}{(1+u)\sqrt{a+bu+cu^2}}\,.
\end{eqnarray}
In the case that we are now considering,
\begin{eqnarray}
a&:=&-\alpha^2 n^2\,,\qquad
b:=\frac{\lambda^2-k^2}{g^2}
  -2  \alpha^2 n^2 r\,,\qquad
c:=-\alpha^2 n^2 r^2 \,,\\
\nonumber\\
 \Delta
&:=&4ac-b^2
 =-\frac{\lambda^2-k^2}{g^2}
  \Big[
  \frac{\lambda^2-k^2}{g^2}
 -4 r \alpha^2 n^2
  \Big]\,.
\end{eqnarray}
Then, we evaluate each term; 
\begin{eqnarray}
&&\left[
 \sqrt{-c}\arcsin
          \left(\frac{2cu+b}{\sqrt{-\Delta}}\right)
-\sqrt{-a}\arcsin
      \left(\frac{2au^{-1}+b}{\sqrt{-\Delta}}\right)
\right]_{u_+}^{u_-}
=-\pi \alpha (1+r) |n|\,,
\end{eqnarray}
and
\begin{eqnarray}
&&-(a-b+c)
 \int_{u_+}^{u_-}
 \frac{du}
      {(1+u)\sqrt{a+bu+cu^2}}
\nonumber\\
&=&\frac{1}{g}
\sqrt{(\lambda^2-k^2)+ g^2 \alpha^2 (1-r)^2 n^2}
\left[
 \arcsin\left(
      \frac{(2a-b)+(b-2c)u_-}
           {(1+u_-)\sqrt{b^2-4ac}}
        \right)
-\arcsin\left(
      \frac{(2a-b)+(b-2c)u_+}
           {(1+u_+)\sqrt{b^2-4ac}}
        \right)        
\right]\,.\label{integral}
\end{eqnarray}

Unfortunately, in order to find the approximated mass spectrum, we must make one further approximation: $\lambda\gg {\rm max}(n,k)$. This limit corresponds to the case where $(b,u_+)\to\infty$ and $u_-\to 0$ respectively. Hence,
\begin{eqnarray}
\int_{y_+}^{y_-}
\sqrt{Q(y)}dy
=\frac{1}{2}
\Big[
-(1+r)\pi \alpha n
+\frac{\pi}{g}
    \sqrt{\lambda^2-k^2+ g^2 \alpha^2 (1-r)^2 n^2}
\Big]
=(m+\frac{1}{2})\pi\,. 
\label{wkbmass}
\end{eqnarray} 
The modes with $n=0$ correspond to the limits $y_{\pm} \to \mp \infty $ and $a,c\to 0$. Thus, from Eq.~(\ref{integral}) the mass spectrum in the WKB approximation becomes {\it exact}:
\begin{eqnarray}
\lambda^2= k^2 + g^2 (2m+1)^2\,.
\end{eqnarray}
For these modes we use a single summation in terms of $m$.

\subsection{KK limit}
To investigate the accuracy of WKB method we can compare the result 
to the non-warped KK limit, where $r=1$. This has an exact analytic solution for the mode functions with
\begin{eqnarray}
  \Phi_{\lambda}(\bar  r)
= A P^{\mu}_{\nu}(\bar r)
 +B P^{-\mu}_{\nu}(\bar r)\qquad\qquad
{z^2}= \frac{1-\bar r}{1+\bar r}\,,
\end{eqnarray}
where we have made a coordinate transformation from $z\to \bar r$ and
\begin{eqnarray}
\nu(\nu+1)= -\frac{1}{4g^2}
    \left(
     k^2-\lambda^2
    \right)\qquad{\rm with} \qquad
\mu=\alpha n\,.
\end{eqnarray}
The exact mass spectrum is then determined by the regularities
of the mode functions on both boundaries,
at ${\bar r}=\pm 1$ and thus, we find 
\begin{eqnarray}
1+\mu+\nu=-m \qquad {\rm or}\qquad \mu-\nu= -m\,,
\end{eqnarray}
where $m=0,1,2,\dots$ as before.
In order to have positivity of the eigenvalue, $\lambda$, we shall take the latter condition. From this the {\it exact} mass spectrum is found to be
\begin{eqnarray}
 \lambda_{m}^2
&=& g^2 (2m+ 2\alpha n)
 (2m+2+2 \alpha  n)
 +k^2
\nonumber\\
&= & g^2 (2m+ 2\alpha n+1)^2
 +k^2
 -g^2
\,.\label{spectrum}
\end{eqnarray}
The result of our WKB method, Eq.~(\ref{wkbmass}), agrees with the one above (when $r=1$) except for 
a constant term in the eigenvalue spectrum. However, given that our WKB approximation is valid for large quantum numbers this constant term is negligible in such a limit. 

\subsection{WKB zeta function}

Given the WKB spectrum derived in the previous subsection, we derive
an expression for the zeta function and the associated effective action.
The zeta function is given by the mode sum
\begin{eqnarray} 
  (2\pi)^2\zeta_{{\rm WKB}} (s,\Delta_{4,\epsilon=0})
=  \int d^2x
\sum_{m,n} \int d^2 k \lambda^{-2s}\,.
\end{eqnarray}
Hereafter, we omit "WKB" in the subscript of the zeta function. 
Here, we shall use the density of states method \cite{Susskind:1994sm}, which 
from Eq. (\ref{wkbmass}) the density of states is easily derived:
\begin{eqnarray}
 \frac{dm}{d\lambda}
=\frac{1}{2g}
 \frac{\lambda}
 {\sqrt{\lambda^2-k^2+n^2g^2 \alpha^2 (1-r)^2}}\,.
\end{eqnarray}
We can therefore replace the sum over $m$ with an integration in terms of
$\lambda$ \cite{Susskind:1994sm}
\begin{eqnarray}
    \sum_m 
    \to
    \int d\lambda\, \frac{dm}{d\lambda}\,.
\end{eqnarray} 
It the following it will be convenient to decompose the zeta function into two pieces, i.e., the $n\neq0$
and $n=0$ modes. 
The total of the derivative of zeta function is finally given by
\begin{eqnarray}
          \zeta{}'(0,\Delta_{4,\epsilon=0})
        = \zeta_{n\neq 0}{}'(0,\Delta_{4,\epsilon=0})
         +\zeta_{n=0}{}'(0,\Delta_{4,\epsilon=0})\,.
\end{eqnarray}
Similar considerations can easily be applied to the six-dimensional case, 
working in the conformal frame.

For the $n \neq 0$ mode, we obtain the following expression
\begin{eqnarray}
 (2\pi)^2\zeta_{n\neq 0 }(s,\Delta_{4,\epsilon=0})
  = \int  d^2x 
  \sum_{n=-\infty}^{\infty}{}'
  \int_{2\sqrt{r}\alpha g n }^{\infty} 
  d\lambda\, \lambda
  \int_0^{\sqrt{\lambda^2-4r \alpha^2 g^2 n^2}}
      dk\, k (2\pi)
       \Big(\frac{dm}{d\lambda}\Big)
       \lambda^{-2s}\,,
\end{eqnarray}
where because of the condition that there be two turning points,
see Eq.~(\ref{conde}), we must restrict the range of integration. The 
$'$ denotes that the $n=0$ mode is to be dropped, because the potential for this mode is not oscillator-like. Performing the $k$-integration first, we obtain
\begin{eqnarray}
 (2\pi)^2\zeta_{n \neq  0}(s,\Delta_{4,\epsilon=0})
&=&\int d^2x \Big\{
 2\sum_{n=1}^{\infty}
  \Big(\frac{\pi}{g}\Big)
\int_0^{\infty} d\tilde\lambda  \,{\tilde \lambda}
\Big(\tilde \lambda^2 
   +4 r \alpha^2 g^2 n^2 
\Big)^{-s}
\sqrt{\tilde\lambda^2 
     + \alpha^2(1+r)^2 g^2 n^2 }  
\nonumber\\
&-& 2\sum_{n=1}^{\infty}
  \Big(\frac{\pi}{g}\Big)
\int_0^{\infty} d\tilde\lambda \,{\tilde \lambda}
\Big(\tilde \lambda^2 
   + 4r \alpha^2 g^2 n^2
\Big)^{-s}
 \big(\alpha g (1+r) n\big)
      \Big\}
      \,,
\end{eqnarray}
where $\tilde \lambda$ is defined by
\begin{eqnarray}
  {\tilde \lambda}^2 
= \lambda^2
 -4 r \alpha^2 g^2 n^2\,.
\end{eqnarray}
For sufficiently large ${\rm Re}(s)$, we obtain the following
integration formulae
\begin{eqnarray}
&&\int_0^{\infty} d{\tilde\lambda}\,{\tilde\lambda}
   \Big(\tilde\lambda^2
   +  4 r \alpha^2 g^2 n^2
  \Big)^{-s}
 =\frac{1}{2(s-1)}
  (4r)^{1-s}
  \big( 
  \alpha g n
  \big)^{2(1-s)}\,,
\\ 
&&\int_0^{\infty} d{\tilde\lambda}\,{\tilde\lambda}
   \Big(\tilde\lambda^2
   +  4r \alpha^2 g^2 n^2
     \Big)^{-s}
  \sqrt{\tilde\lambda^2 
     +\alpha^2 g^2 (1+r)^2 n^2}  
 \nonumber \\
 &=&(\alpha g n)^{3-2s}
 \Big\{
 -\frac{\Gamma(1-s)\Gamma(s-\frac{1}{2})}
       {2(2s-3)\sqrt{\pi}}
  (1-r)^{3-2s}
 +\frac{1}{2(s-1)}
 (4r)^{1-s}
 (1+r)
 {}_2 F_{1}[-\frac{1}{2}, 1,2-s,
           \frac{4r}{(1+r)^2}] 
 \Big\} \,,  \label{yamada}
\end{eqnarray}
where again ${}_2F_1(a,b,c;z)$ is a hypergeometric function. Note that 
the first term in Eq.~(\ref{yamada}) has been analytically continued by employing the standard duplication formula for the Gamma function; otherwise the result will be ill defined when taking the derivative at $s=0$.

Finally, expressing the summations over $n$ in terms Riemann zeta functions we obtain (substituting back the expression for $\alpha$, see Eq. (C6)) 
\begin{eqnarray}
 (2\pi)^2 \zeta_{ n \neq 0}(s,\Delta_{4,\epsilon=0})
&=&\int d^2 x
\Big[ \frac{2\pi}{g}  (\alpha g)^{3-2s} 
 \zeta_{R}(2s-3)
 \Big(\frac{2\pi}{2\pi-\delta_+}\Big)^{3-2s}
  \Big\{
  -\frac{\Gamma(1-s)\Gamma(s-\frac{1}{2})}
        {2(2s-3)\sqrt{\pi}}(1-r)^{3-2s}
 \nonumber\\ 
 && +\frac{(4r)^{1-s}(1+r)}{2(s-1)}
   \Big(
     {}_2 F_{1}[-\frac{1}{2}, 1,2-s;
           \frac{4r}{(1+r)^2}] 
     -1
   \Big)
  \Big\}\Big]\,,
\end{eqnarray}
where $\zeta_R(s)$ is Riemann's zeta function. Then, taking the derivative in terms of $s$ and an analytic continuation $s \to 0$, we find
\begin{eqnarray}
 \zeta_{ n \neq 0} '(0,\Delta_{4,\epsilon=0})
&=&  \frac{g^2 \pi^2}{270(2\pi-\delta_+)^3}
\int\, d^2x
 \Big\{ \big(-2-3\gamma -3\psi(-\frac{1}{2})\big)(1-r)^3
 +2(-3+r)r^2 
             \Big(
             -3 + 3\ln \Big(\frac{4r}{(1-r)^2}\Big)
             \Big)
\nonumber\\
&+& (1+r) (1-4r+r^2)
      \Big(6 \ln\big(\frac{2\pi g (1-r)}{2\pi-\delta_+}\big) 
          -720 \zeta_R'(-3)
      \Big)
   +18(1+r)r \frac{\partial}{\partial c}
     {}_2F_1[-\frac{1}{2},1,2,\frac{4r}{(1+r)^2}]
  \Big\}     
     \,.
\end{eqnarray} 

The zeta function for $n=0$ mode is given by
\begin{eqnarray}
 (2\pi)^2\zeta_{n=0} (s,\Delta_{4,\epsilon=0})
&=& \int d^2x
    \int d^2k 
    \sum_{m=0}^{\infty}
    \big[g^2 (2m+1)^2+k^2 \big]^{-s}
= \frac{\pi (2g)^{2-2s}}{(s-1)}
  \zeta_{H} (2s-2,\frac{1}{2})
\,,
\end{eqnarray}
where $\zeta_{H}(s,a)$ is Hurwitz's zeta function.
Taking the derivative with respect to $s$ and continuing to $s\to 0$ gives
\begin{eqnarray}
 \zeta_{n=0}{}'(0,\Delta_{4,\epsilon=0})
=\frac{3g^2\zeta_{R}'(-2)}{2\pi}  \,.
\end{eqnarray}

The total result of Eq. (\ref{4dcoef}) or Eq. (\ref{a_tot}) is 
\begin{eqnarray}
&& A_{4 , {\rm WKB}}(r,\delta_+)
= -\frac{g^2 \pi^2}{540(2\pi-\delta_+)^3}
 \Big\{ \big(-2-3\gamma -3\psi(-\frac{1}{2})\big)(1-r)^3
        +2(-3+r)r^2 
             \Big(
             -3 + 3\ln \Big(\frac{4r}{(1-r)^2}\Big)
             \Big)\,
\\
&+& (1+r) (1-4r+r^2)
      \Big(6 \ln\big(\frac{2\pi g (1-r)}{2\pi-\delta_+}\big) 
          -720 \zeta_R'(-3)
      \Big)
 +18(1+r)r
 \frac{\partial}{\partial c}
     {}_2F_1[-\frac{1}{2},1,2,\frac{4r}{(1+r)^2}]
  \Big\}     
+\frac{3g^2 \zeta_{\rm R} {}'(-2)}{4\pi}
     \,.
     \nonumber
     \label{inflationary3}
\end{eqnarray}



\end{document}